\documentclass[article,twocolumn,
superscriptaddress,
amsmath,amssymb,
aps,
pra,
floatfix,
]{revtex4-2}

\usepackage[normalem]{ulem}
\usepackage{graphicx}
\usepackage{xfrac}
\usepackage{amsmath}
\usepackage{mathtools}

\usepackage{epsfig}
\usepackage[colorlinks=true,citecolor=blue,linkcolor=blue,urlcolor=blue]{hyperref}
\usepackage{MnSymbol}
\usepackage{siunitx}
\usepackage{float}

\newcommand{\Red}[1]{{\color{red}{#1}}}

\usepackage{graphicx}% Include figure files
\usepackage{dcolumn}% Align table columns on decimal point
\usepackage{bm}% bold math
\usepackage{tikz,xcolor,hyperref}

\definecolor{lime}{HTML}{A6CE39}
\DeclareRobustCommand{\orcidicon}{
	\begin{tikzpicture}
	\draw[lime, fill=lime] (0,0) 
	circle [radius=0.16] 
	node[white] {{\fontfamily{qag}\selectfont \tiny ID}};
	\draw[white, fill=white] (-0.0625,0.095) 
	circle [radius=0.007];
	\end{tikzpicture}
	\hspace{-2mm}
}
\foreach \x in {A, ..., Z}{\expandafter\xdef\csname orcid\x\endcsname{\noexpand\href{https://orcid.org/\csname orcidauthor\x\endcsname}
			{\noexpand\orcidicon}}
}
\begin{document}

\title{Narrow and ultra-narrow transitions in highly charged Xe ions as probes of fifth forces}
	
\author{Nils-Holger Rehbehn\orcidA{}}
\email[]{nils.rehbehn@mpi-hd.mpg.de}
\affiliation{Max-Planck-Institut f\"ur Kernphysik, D--69117 Heidelberg, Germany}

\author{Michael K. Rosner\orcidB{}}
\affiliation{Max-Planck-Institut f\"ur Kernphysik, D--69117 Heidelberg, Germany}

\author{Julian C. Berengut\orcidC{}}
\affiliation{School of Physics, University of New South Wales, Sydney, New South Wales 2052, Australia}
\affiliation{Max-Planck-Institut f\"ur Kernphysik, D--69117 Heidelberg, Germany}

\author{Piet O.~Schmidt\orcidD{}}
\affiliation{Physikalisch--Technische Bundesanstalt, D--38116 Braunschweig, Germany}
\affiliation{Leibniz Universit\"at Hannover, D--30167 Hannover, Germany}

\author{Thomas Pfeifer\orcidL{}}
\affiliation{Max-Planck-Institut f\"ur Kernphysik, D--69117 Heidelberg, Germany}

\author{Ming Feng Gu}
\affiliation{Space Science Laboratory, University of California, Berkeley, CA 94720, USA}

\author{Jos\'e R. {Crespo L\'opez-Urrutia}\orcidJ{}}
\email[]{crespojr@mpi-hd.mpg.de}
\affiliation{Max-Planck-Institut f\"ur Kernphysik, D--69117 Heidelberg, Germany}

\date{\today}
	
\begin{abstract}
Optical frequency metrology in atoms and ions can probe hypothetical fifth-forces between electrons and neutrons by sensing minute perturbations of the electronic wave function induced by them. A generalized King plot has been proposed to distinguish them from possible Standard Model effects arising from, e.g., finite nuclear size and electronic correlations. Additional isotopes and transitions are required for this approach. Xenon is an excellent candidate, with seven stable isotopes with zero nuclear spin, however it has no known visible ground-state transitions for high resolution spectroscopy. To address this, we have found and measured twelve magnetic-dipole lines in its highly charged ions and theoretically studied their sensitivity to fifth-forces as well as the suppression of spurious higher-order Standard Model effects. Moreover, we identified at 764.8753(16)\,nm a E2-type ground-state transition with 500\,s excited state lifetime as a potential clock candidate further enhancing our proposed scheme.
\end{abstract}
	
\maketitle
	
%\section{\label{sct:introduction}Introduction}
Indirect evidence from galactic rotation, gravitational lensing and cosmological evolution suggests the existence of dark matter (DM) \cite{Bertone_2005_Review, bertone_history_2018-1}. Its constituents, by coupling to Standard Model (SM) particles, could also influence the neutrino-mass hierarchy \cite{Alvey_2019} and explain open physics questions. Additional fields could cause a fifth-force coupling of electrons with neutrons, inducing small but measurable effects \cite{Safronova2018, Frugiuele2017, Berengut2018} in atomic systems. 
In an electronic transition, sensitivity to a fifth-force arises because the overlap with the nucleus changes between the ground and excited state, reflecting interactions as small as a fraction of the linewidth.

However, since the transition energies cannot be calculated accurately, isotope shift spectroscopy is employed. The classical method of the King plot (KP) uses two transitions in at least three different isotope pairs are used. Plotting the isotope shifts scaled by the nuclear-mass parameter eliminates the charge radius, typically leading to a linear behavior~\cite{king_1963} from which atomic constants characterising atomic recoil (mass shift) and the overlap of the electronic wavefunction with the nucleus (field shift) can be derived. Optical frequency metrology has reduced uncertainties in the determination of transition energies of forbidden transitions by orders of magnitude, making KP methods far more sensitive.
On top of such isotopic shifts (IS), hypothetical fifth forces would add minute perturbations causing a deviation from linearity \cite{Berengut2018,Frugiuele2017,Flambaum2018,Fichet2018,Drake2021,Dhindsa2022}. However, unknown SM effects of higher order \cite{Berengut2018,Flambaum2018} (sometimes dubbed `spurions') could also induce sizable non-linearities that are hard to distinguish from those of the hypothetical forces.

A recently devised generalized King plot (GKP) \cite{Mikami2017, berengut_2020} overcomes this by using more transitions and isotope pairs to build a set of linear equations determining the higher-order effects, and even disposing of the need of exact nuclear masses (`no-mass GKP'). Recent dedicated experiments in ytterbium measured a deviation from the King linearity~\cite{counts_observation_2020}, which has since been confirmed using other transitions~\cite{Figueroa_2022,ono22prx,hur22prl}. The most likely reason for the deviation is changes in the nuclear deformation of Yb between isotopes~\cite{Allehabi_2021}, however the analysis show that a second, yet unidentified source of nonlinearity is also present.
In contrast, a measurement in calcium \cite{Solaro2020}, where higher-order effects are expected to be smaller, was consistent with King linearity. 

To further enhance the GKP sensitivity, as many transitions and even-isotopes as possible are sought after. An ideal candidate is xenon ($Z=54$). It has seven natural zero-nuclear-spin isotopes (124, 126, 128, 130, 132, 134, 136), and four more radio-isotopes (118, 120, 122 and 138) with lifetimes longer than minutes. 
Its mass reduces Doppler shifts in comparison with lighter elements, and its nuclear charge magnifies relativistic and QED effects \cite{Shabaev2018}, which is in itself a key field of research \cite{Indelicato_2019}. 

\begin{figure*}%%% placed here due to stuck figure
\centering
\includegraphics[width=\linewidth]{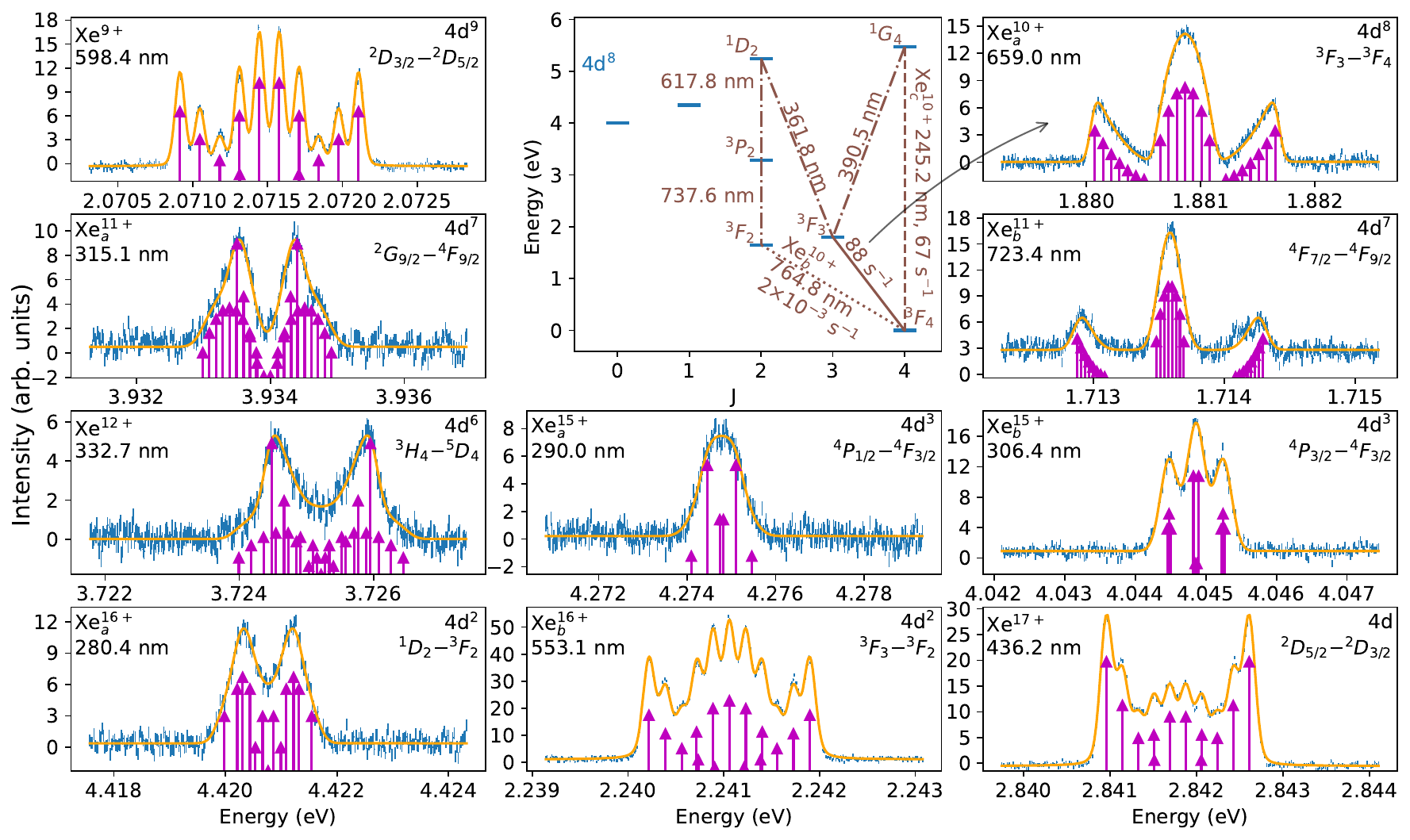}
\caption{Measured optical ground-state transitions of Xe$^{9+}$ through Xe$^{17+}$. The Zeeman structure (arrows) was fitted based on line identification with \textsc{fac} calculations. For Xe$^{10+}$ the Grotrian diagram has been expanded. Dotted line: calculated E2 transition; dashed line: calculated M1 transition (see Supplemental Materials [\Red{URL will be inserted by publisher}] for details and further Grotrian diagrams).}
\label{fig:Xe_overview}
\end{figure*}

In this Letter, we find twelve xenon ground-state, magnetic dipole (M1) transitions in the optical region, with wavelength uncertainties in the order of $0.1\,\mathrm{pm}$, and one optical electric quadrupole (E2) clock transition in charge states Xe$^{9+}$ through Xe$^{17+}$. We have calculated and evaluated theoretical King plots to find which combination of transitions leads to the highest possible sensitivity to a hypothetical fifth-force between neutrons and electrons.

Highly charged ions (HCI) such as those studied here are very well suited for high-precision experiments \cite{KozlovRMP2018}, since their very low polarizability suppresses effects of external electromagnetic perturbations. Moreover, the reduced number of bound electrons reduces their theoretical complexity. Recent experimental developments like sympathetic cooling \cite{schmoger_coulomb_2015}, application of quantum logic spectroscopy \cite{schmidt_2005} to HCI \cite{micke_2020}, and algorithmic cooling \cite{king_algorithmic_2021} have made clocks based on HCI  with a sub-Hz uncertainty possible \cite{King2022}. They will help extending GKP applications into beyond-the-SM (BSM) parameter regions not yet constrained by scattering experiments \cite{barbieri_1975,leeb_1992,nesvizhevsky_2008,pokotilovski_2006,adler_1974} and fifth-force studies \cite{bordag_2001,bordag_2009,germann_three-body_2021, salumbides_bounds_2013, jaffe_testing_2017, biesheuvel_probing_2016}.

%\section{\label{sct:experiment}Experiment}

\begin{table*}
\centering
\caption{Transitions in highly charged Xe ions: Measured energies, wavelengths ($\lambda_\mathrm{vac}$, vacuum) and g-factors of upper and lower energy levels obtained through fitting of the Zeeman-structure with their corresponding uncertainties. Theoretical transition probabilities $A_{ki}$, ab-intio wavelengths and g-factors were calculated with \textsc{ambit} \cite{ambit_2019}.}
\begin{tabular}{lllllllllrrrrrr}
\hline
\hline
& \multicolumn{4}{c}{}&	\multicolumn{4}{c}{Observed values} & \multicolumn{5}{c}{\textsc{ambit} calculations} &  \\
Ion & \multicolumn{4}{c}{Transition} &	Energy (eV) & $\lambda_\mathrm{vac}$ (nm) & $g_\mathrm{upper}$ & $g_\mathrm{lower}$ & $A_{ki}$ (s$^{-1}$) & $\lambda$ (nm) & $g_\mathrm{upper}$ & $g_\mathrm{lower}$ & Type\\

\hline
&&&&&&&&&& \\
Xe$^{9+}$&$4d^9$&$ ^2\!D_{3/2} $&-&$ ^2\!\!D_{5/2}$ & 2.07151156(28) & 598.520419(90) & $0.792(2)$ & $1.189(1)$ & 67.5 & 595.4 & 0.8 & 1.2 & M1 \\

Xe$^{10+}_a$&$4d^8$&$ ^3\!F_3 $&-&$ ^3\!\!F_4$ & 1.8808627(14) & 659.18789(56) & $1.082(4)$ & $1.238(3)$ & 88.4 & 665.1 & 1.0833 &1.2426 & M1 \\ 
Xe$^{10+}_b$&$4d^8$&$ ^3\!F_2 $&-&$ ^3\!\!F_4$ &  1.6209726(34) & 764.8753(16)(Ritz) &&& 0.002 &735.6&0.9792&1.2426& \textbf{E2} \\
Xe$^{10+}_c$&$4d^8$&$ ^1\!G_4 $&-&$ ^3\!\!F_4$ & 5.0567231(73) & 245.18684(36)(Ritz) &&& 67.1 &242.9&1.0074&1.2426& M1 \\

Xe$^{11+}_a$&$4d^7$&$ ^2\!G_{9/2} $&-&$ ^4\!\!F_{9/2}$  & 3.9339517(50) & 315.16451(45) & $1.105(37)$ & $1.321(35)$ & 107.7 & 313.3 & 1.0823 & 1.3054 & M1 \\  
Xe$^{11+}_b$&$4d^7$&$ ^4\!F_{7/2} $&-&$ ^4\!\!F_{9/2}$  & 1.71358555(87) & 723.53667(37) & $1.242(9)$ & $1.306(7)$ & 82.8 & 727.4 & 1.2276 & 1.3054 & M1 \\  

Xe$^{12+}$&$4d^6$&$ ^3\!H_4 $&-&$ ^5\!\!D_4$ & 3.7252183(78) & 332.82398(79) & $1.058(32)$ & $1.455(31)$ & 110.8 & 328.6 & 1.0438 & 1.462 & M1 \\  

Xe$^{15+}_a$&$4d^3$&$ ^4\!P_{1/2} $&-&$ ^4\!\!F_{3/2}$ & 4.2747781(74) & 290.03657(50) & $2.17(20)$ & $0.78(9)$ & 16.8 & 297.7 & 2.2137 & 0.47304 & M1 \\ 
Xe$^{15+}_b$&$4d^3$&$ ^4\!P_{3/2} $&-&$ ^4\!\!F_{3/2}$ & 4.0448609(24) & 306.52277(20) & $0.856(57)$ & $0.799(28)$ & 101.7 & 337.1 & 1.4357 & 0.47304 & M1 \\ 

Xe$^{16+}_a$&$4d^2$&$ ^1\!D_2 $&-&$ ^3\!\!F_2$& 4.4207651(39) & 280.45869(25) & $1.197(11)$ & $0.704(15)$ & 84.4 & 271.3 & 1.2027 & 0.70681 & M1 \\ 
Xe$^{16+}_b$&$4d^2$&$ ^3\!F_3 $&-&$ ^3\!\!F_2$& 2.24105833(38) & 553.23949(11) & $1.073(1)$ & $0.702(2)$& 137.6 & 555.4 & 1.0833 & 0.70681 & M1 \\ 
Xe$^{16+}_c$&$4d^2$&$ ^3\!P_1 $&-&$ ^3\!\!F_2$ & 4.9243614(65) & 251.77721(33)(Ritz) &&& 10.3 &244.3& 1.5 & 0.70681 & M1 \\

Xe$^{17+}$&$4d$&$ ^2\!D_{5/2} $&-&$ ^2\!\!D_{3/2}$& 2.84178291(47) & 436.290179(73) & $1.184(2)$ & $0.785(3)$ & 123.7 & 433.8 & 1.2 & 0.8 & M1 \\
&&&&&&&&&&&&&& \\
\hline
\hline
\end{tabular}
\label{tab:results}
\end{table*}

We produced the ions of interest with an electron beam ion trap (EBIT) \cite{levine_electron_1988,levine_use_1989}, the Heidelberg EBIT (HD-EBIT) \cite{crespo_1999a}, from a differentially pumped atomic beam of Xe interacting with electrons at selected energies. For producing the HCI of interest Xe$^{9+}$ through Xe$^{33+}$, we scanned the electron beam energy in the range 100-2500\,eV in 10-eV increments. Every time the ionization potential of a given charge state is surpassed, the next higher one appears in the trap, and with it a different spectrum. In each cycle, we kept these ions trapped for 60\,s, and dumped them at the end by briefly inverting the axial trapping potential set by trap electrodes. This removes impurity ions that slowly accumulate due to evaporation of barium and tungsten from the electron-gun cathode.  

\begin{figure} %%% moved up here for better page layout
\centering
\includegraphics[width=\linewidth]{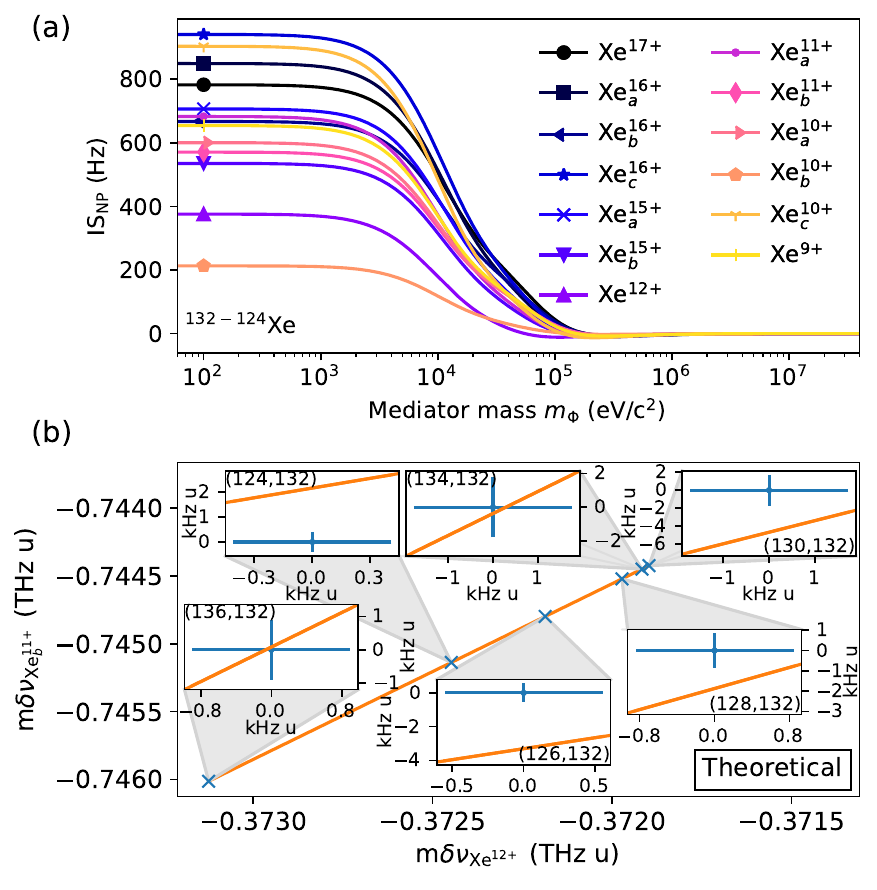}
\caption{(a) Isotopic shift from a hypothetical fifth-force for the forbidden ground-state transitions in Xe$^{9+}$ through Xe$^{17+}$ for a fixed coupling constant $y_ey_n=1\times10^{-13}$ and varying mediator mass.
(b) Theoretical King plot for the six possible Xe-isotope pairs using the Xe$^{11+}_b$, Xe$^{12+}$ pair as example. We set the coupling constant to $y_ey_n=1\times10^{-13}$ and the mediator mass to $m_\Phi=1\times10^5\,\frac{\mathrm{eV}}{\mathrm{c}^2}$. The error bars represent 100~mHz measurement uncertainty modified by the mass parameter $\mu_a$.}
\label{fig:Xe_IS_and_KP}
\end{figure}

Electron-impact excitation populates the upper levels of the observed transitions both directly and through cascades.
The electron and ion density conditions in the EBIT are such that optical magnetic-dipole transitions with Einstein coefficients higher than $\approx$ 10\,s$^{-1}$ can be measured \cite{Draganic2003,SoriaOrts2006,Bekker2018,Bekker2019,Rehbehn_2021}. %with good signal-to-noise-ratio (SNR). 
For this purpose, a set of in-vacuo lenses projects an image of the cylindrical ion cloud through a quartz vacuum window. This intermediate image is rotated by 90 degrees by a periscope and relayed by two lenses to the entrance slit of an optical spectrometer, as in Refs. \cite{Draganic2003,SoriaOrts2006}. In the present work, we used a Czerny-Turner spectrometer with 2\,m focal length \cite{Bekker2018,Bekker2019,Rehbehn_2021} to record the wavelength range 250-800\,nm, and calibrated it with hollow-cathode lamps of different elements. In its focal plane, a CCD-camera, cooled to $-80^\circ\mathrm{C}$, took several images for averaging with an exposure time of 60 minutes each. Pixels showing high signals due to cosmic muons were identified and removed from the images. Stray-light background was also subtracted. After obtaining overview spectra with a 150 grooves/mm grating, we performed measurements at higher resolution using two gratings with 1800 and 3600 grooves/mm, respectively. The results are shown in Figure \ref{fig:Xe_overview}. 

For identification, we calculated for each ion the electronic structure and transition rates with the Flexible Atomic Code  (\textsc{fac}) \cite{gu_2008} and \textsc{ambit} \cite{ambit_2019}. The advantage of \textsc{fac} is its calculation speed, while \textsc{ambit} is used for its more precise results. Since the 8-T field of the EBIT separates the Zeeman components to a resolvable extent, we fitted for each line its centroid, experimental linewidth, the g-factors of the upper and lower state, $\pi$ and $\sigma$ amplitudes, and compare the results with theory. %For this, we use wavefunctions generated by \textsc{fac} to infer LS-Terms with the \textsc{grasp} \textsc{jj2lsj} function \cite{JONSSON_2013}, and obtain g factors and their corresponding Clebsch-Gordan coefficients. 
We corrected the identification of the 436.2\,nm line in Ref. \cite{crespo_1999b}, from Xe$^{18+}$ to Xe$^{17+}$ based on the Zeeman splitting. Table \ref{tab:results} presents the key parameters of the thirteen discovered ground-state transitions. Wavelength uncertainties are the quadratic average of those from the Zeeman fits and the spectrometer dispersion. Results of \textsc{ambit} calculations are also given there: ab-initio wavelength $\lambda$, Einstein coefficient $A_{ki}$, and expected g-factors. Furthermore, we tabulate in the Supplemental Material [\Red{URL will be inserted by publisher}] many other identified lines not involving the electronic ground-state.

A key result of our search is the identification of a ground-state clock transition of E2 electric quadrupole in Xe$^{10+}$ from the lowest excited state. By means of Ritz-Rydberg combinations (see Supplemental Materials [\Red{URL will be inserted by publisher}]), we determine for the $4d^8$ $ ^3\!F_2 \rightarrow ^3\!\!\!F_4$ transition a vacuum wavelength of 764.8753(16)\,nm. We compare this with an \textsc{ambit} calculation yielding 735.6\,nm and an E2 transition rate of $A_{ki}=0.002\,\mathrm{s}^{-1}$, which is within the expected calculation uncertainty. The low transition rate suggests that this is a suitable candidate for an optical clock with a sensitivity to fifth forces that can be fully exploited due to its narrow linewidth of 0.3\,mHz. 
We also have preliminary assignments for a few additional E2 candidates in other charge states in the Supplemental Materials [\Red{URL will be inserted by publisher}], but a conclusive identification will require complementary measurements.

% \section{\label{sct:predictions}Theoretical predictions} 
% \begin{figure*} %%% bigger exclusion plot image, might not fit within the page limit of PRL
% \centering
% \includegraphics[width=\linewidth]{Xe_exclusion_plotter_figure_wide.pdf}
% \caption{Lower limit of the coupling constant $y_\mathrm{e}y_\mathrm{n}$ for selected transition pairs with assumed frequency uncertainty of 100 mHz. Dashed lines: Classical King plot (KP) neglecting mass uncertainty. Calcium King plot from \citet{Solaro2020} with 100~mHz uncertainty included for comparison. Dash-dotted line: Inclusion of a second-order field shift of $1\,\mathrm{kHz}$ and mass uncertainty. Solid line: Calculation of the no-mass generalized King plot (GKP) \cite{berengut_2020} using four transitions; in comparison with GKP Yb prediction from \citet{berengut_2020}.}
% \label{fig:Xe_exclusion}
% \end{figure*}

Following an established approach, we checked the suitability of the found ground-state transitions for GKP studies. For this, we added to the electromagnetic interaction potentials used for the \textsc{fac} calculations an additional term for the hypothetical fifth-force as a Yukawa potential \cite{Mikami2017}:
\begin{equation}
    V_\Phi(r)=y_\mathrm{e}y_\mathrm{n}(A-Z)\frac{\hbar c}{4\pi r}\exp\left(-\frac{c}{\hbar}m_\Phi r\right),
    \label{eq:Vphi}
\end{equation}
where $y_\mathrm{e}y_\mathrm{n}$ is the coupling strength between electrons and neutrons; $A$ the nuclear mass, $Z$ the nuclear charge; $\hbar$ the reduced Planck constant, and $c$ the speed of light. The range of the force is defined by the mass parameter $m_\Phi$. 
Using an automatized script, we performed \textsc{fac} calculations varying the force range, its strength, as well as the nuclear charge radius and mass, and calculated the corresponding perturbation on the isotope-shifts. 

Between two isotopes, the total isotope shift (IS) $\delta\nu_i^a=\nu_{A_\mathrm{ref.}}-\nu_A$ is expressed by the following equation \cite{Mikami2017} for the transition $i$ and the isotope pair $a=(A_\mathrm{ref.},A)$: 
\begin{equation}
    \delta\nu_i^a=K_i\mu_a+F_i\delta\langle r^2 \rangle_a +y_\mathrm{e}y_\mathrm{n}X_i\gamma_a.
    \label{eq:IS}
\end{equation}
Here, the first two terms represent the mass shift and the field shift, respectively. The former depends on the difference of the inverse of the isotope masses $\mu_a=1/m_{A_\mathrm{ref.}}-1/m_A$, and the latter on the mean-square charge radii difference $\delta\langle r^2 \rangle_a=\langle r^2 \rangle_{A_\mathrm{ref.}}-\langle r^2 \rangle_A$. 
The third term is caused by the fifth force, and depends on the coupling strength $y_\mathrm{e}y_\mathrm{n}$ between electrons and neutrons, the electronic constant $X_i=X_i(V_\Phi)$ due to the Yukawa potential, and a factor $\gamma_a=(A_\mathrm{ref.}-Z)-(A-Z)$ depending on the difference in number of neutrons. We plot the fifth-force shift versus varying mediator masses $m_\Phi$ in Fig.~\ref{fig:Xe_IS_and_KP}(a). 

To study the effect of a fifth-force, we used a King plot \cite{king_1963}, where the isotope shift in Eq.~\ref{eq:IS} is divided by the mass parameter $\mu_a$, which then yields a common nuclear parameter $\delta\langle r^2 \rangle_a/\mu_a$ in the SM part. Between two modified transitions, the SM parts lead to a linear behavior, but a fifth-force would break it. For the six Xe-isotope pairs this is shown as theoretical prediction in Fig.~\ref{fig:Xe_IS_and_KP}(b). The mediator mass $m_\Phi=1 \times 10^5 \frac{\mathrm{eV}}{\mathrm{c}^2}$ and the coupling parameter $y_\mathrm{e}y_\mathrm{n}=10^{-13}$ are fixed at arbitrary, but theoretically possible values \cite{Berengut2018}. Mass uncertainties are neglected. The fifth-force shift in Fig.~\ref{fig:Xe_IS_and_KP}(a) is on the order of hundreds of Hz, but only a small fraction remains as a non-linearity due to the alignment of the SM linearity with the fifth-force contributions \cite{Berengut2018}. 

\begin{figure} %%% smaller exclusion plot image
\centering
\includegraphics[width=\linewidth]{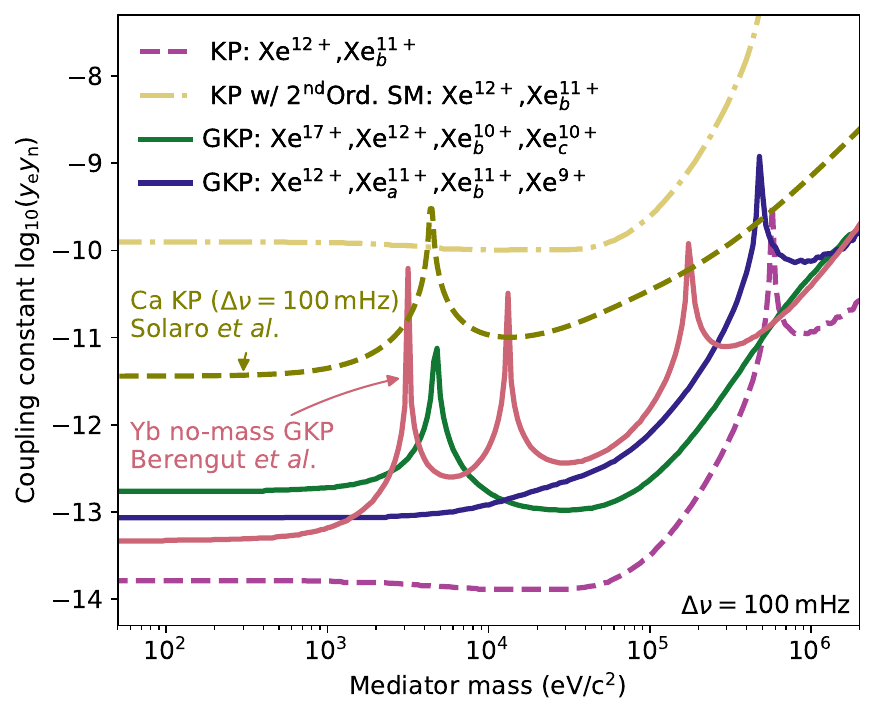}
\caption{Sensitivity limit for $y_\mathrm{e}y_\mathrm{n}$ for selected transition pairs with assumed frequency uncertainty of 100 mHz. Dashed lines: Classical King plot (KP) neglecting mass uncertainty. Calcium King plot from \citet{Solaro2020} with 100~mHz uncertainty included for comparison. Dash-dotted line: Inclusion of a second-order field shift of $1\,\mathrm{kHz}$ and mass uncertainty. Solid lines: Predicted no-mass GKP for Yb \cite{berengut_2020} and for the present Xe, using four transitions.}
\label{fig:Xe_exclusion}
\end{figure}

Although each pairing of transitions experiences a different non-linearity, similar fine-structure transitions share common-mode shifts, reducing the total sensitivity as in Refs.~\cite{berengut_2020,Rehbehn_2021}. Following these works, we quantified non-linearities in the King plot with the area $\mathrm{NL}$ spanned by the isotope pairs. Error propagation of the assumed measurement uncertainty on the isotope-shifts yielded its uncertainty $\Delta\mathrm{NL}$. With this, we defined the resolution $R$ as $R=\mathrm{NL}/\Delta\mathrm{NL}$.
A value $R=1$ sets the lower bound of the coupling parameter $y_\mathrm{e}y_\mathrm{n}$ that can be resolved, as shown for given transition pairs in Fig.~\ref{fig:Xe_exclusion}. 

We assume a frequency uncertainty of $\Delta\nu=100\,\mathrm{mHz}$ as recently achieved \cite{micke_2020, King2022} and discussed in Ref.~\cite{Rehbehn_2021}. %The numerical accuracy of our \textsc{fac} calculations is insufficient for distinguishing effects of fifth forces from nuclear ones for mediator masses above $2 \times 10^6 \,\mathrm{eV}/\mathrm{c}^2$.
Dashed lines in Fig.~\ref{fig:Xe_exclusion} depict King plots neglecting mass uncertainty and higher-order SM effects. Inclusion of the former in one of these pairs, e.g., Xe$^{12+}$, Xe$^{11+}_{b}$, for isotopes 130 to 136 (relative mass uncertainties $\approx 10^{-10}$) and 124 to 128 (between $10^{-8}$ and $10^{-7}$) \cite{xenon_weight} diminishes sensitivity by three orders of magnitude. Another order of magnitude is lost if nuclear deformation causes a second-order field shift of $1\,\mathrm{kHz}$ (dash-dotted line) \cite{Allehabi_2020,Allehabi_2021} estimated by evaluating calculated higher-order shifts in other elements \cite{Flambaum2018}. 
We can overcome these losses with a no-mass generalized King plot \cite{berengut_2020} (shown as solid line) by adding more transitions to expand the King plot into a higher dimension. This restores the sensitivity by three orders of magnitude, as shown in Fig.~\ref{fig:Xe_exclusion} and brings it to the level of the ytterbium GKP with an assumed frequency uncertainty of $100\,\mathrm{mHz}$ from Ref.~\cite{berengut_2020}. A different xenon pairing would improve the sensitivity around $10^5$~eV.
Full sensitivity is not recovered due to the error propagation of the four transitions needed. 
Note that either a reduction of the $\nu$-uncertainty down to 5~mHz, as already achieved by optical clocks \cite{Bloom2014}, or a separate, sufficiently accurate mass measurement would lead to better sensitivity.

If more higher-order SM contributions are expected, more transitions can be used to suppress their spurious effects. The seven stable even isotopes of Xe can tackle up to three such `spurions' by expanding the KP into five dimensions. With thirteen ground-state transitions, one can use various pairings to optimize sensitivity to NP. By contrast, other candidates such as calcium or ytterbium only have enough even isotopes to suppress one `spurion'.

% \section{\label{sct:discussion_conclusion}Discussion and conclusion}
In the future, we plan to extend our search towards the EUV region to include ground-state transitions of even higher-charged xenon ions. This increases the number of transitions and their GKP sensitivity, as electronic wave functions with stronger overlap with the nucleus are involved. Currently emerging and continuously developing EUV frequency combs \cite{Cingoez_2012,Lyu_2020,Nauta_2017,Nauta_2021} should in the future allow frequency metrology on HCI in this spectral regime. Moreover, the presently assumed uncertainty of $100\,\mathrm{mHz}$ is due to that of the SI second reference, which is a Cs transition \cite{Weyers_2018,guena_first_2017}. By performing direct frequency comparisons with optical clocks \cite{Rosenband2008,Godun2014,Beloy2020}, or by measuring the isotope shift of simultaneously trapped ions \cite{Manovitz2019}, one could enhance sensitivity even by a few orders of magnitude.

In summary, we found and identified thirteen optical ground-state transitions in highly charged xenon ions, theoretically analyzed them, and evaluated pairs with high sensitivity to Yukawa-type fifth-forces. Such pairings overcome limitations caused by both known and unknown SM effects by enabling the method of no-mass generalized King plots. We found for xenon a sensitivity already comparable to ytterbium, but with the clear advantage of its more numerous even isotopes and reduced nuclear deformation for future frequency metrology with sub-Hz uncertainty. Moreover, we identified a ground-state ultra-narrow clock (E2) transition with 0.3\,mHz linewidth complementing the other twelve M1 lines. The analysis of this set of transitions can also be used to conduct further tests of nuclear deformation \cite{Zyriliou_2022}. 

\section*{Acknowledgements}
Financial support was provided by the Max-Planck-Gesellschaft. We acknowledge support from the Max Planck-Riken-PTB Center for Time, Constants and Fundamental Symmetries, the collaborative research center “SFB 1225 (ISOQUANT)” and the Deutsche Forschungsgemeinschaft (DFG, German Research Foundation) under Germany’s Excellence Strategy – EXC-2123 QuantumFrontiers – 390837967. JCB was supported in this work by the Alexander von Humboldt Foundation and the Australian Research Council (DP190100974). This work has been funded by the “European Metrology Program for Innovation and Research” (EMPIR) project 20FUN01 TSCAC. This project has received funding from the EMPIR programme co-financed by the participating states and from the European Union’s Horizon 2020 research and innovation programme. This project has received funding from the European Research Council (ERC) under the European Union’s Horizon 2020 research and innovation programme (grant agreement No 101019987).

\bibliography{literature.bib}

%apsrev4-2.bst 2019-01-14 (MD) hand-edited version of apsrev4-1.bst
%Control: key (0)
%Control: author (8) initials jnrlst
%Control: editor formatted (1) identically to author
%Control: production of article title (0) allowed
%Control: page (0) single
%Control: year (1) truncated
%Control: production of eprint (0) enabled
\begin{thebibliography}{63}%
\makeatletter
\providecommand \@ifxundefined [1]{%
 \@ifx{#1\undefined}
}%
\providecommand \@ifnum [1]{%
 \ifnum #1\expandafter \@firstoftwo
 \else \expandafter \@secondoftwo
 \fi
}%
\providecommand \@ifx [1]{%
 \ifx #1\expandafter \@firstoftwo
 \else \expandafter \@secondoftwo
 \fi
}%
\providecommand \natexlab [1]{#1}%
\providecommand \enquote  [1]{``#1''}%
\providecommand \bibnamefont  [1]{#1}%
\providecommand \bibfnamefont [1]{#1}%
\providecommand \citenamefont [1]{#1}%
\providecommand \href@noop [0]{\@secondoftwo}%
\providecommand \href [0]{\begingroup \@sanitize@url \@href}%
\providecommand \@href[1]{\@@startlink{#1}\@@href}%
\providecommand \@@href[1]{\endgroup#1\@@endlink}%
\providecommand \@sanitize@url [0]{\catcode `\\12\catcode `\$12\catcode
  `\&12\catcode `\#12\catcode `\^12\catcode `\_12\catcode `\%12\relax}%
\providecommand \@@startlink[1]{}%
\providecommand \@@endlink[0]{}%
\providecommand \url  [0]{\begingroup\@sanitize@url \@url }%
\providecommand \@url [1]{\endgroup\@href {#1}{\urlprefix }}%
\providecommand \urlprefix  [0]{URL }%
\providecommand \Eprint [0]{\href }%
\providecommand \doibase [0]{https://doi.org/}%
\providecommand \selectlanguage [0]{\@gobble}%
\providecommand \bibinfo  [0]{\@secondoftwo}%
\providecommand \bibfield  [0]{\@secondoftwo}%
\providecommand \translation [1]{[#1]}%
\providecommand \BibitemOpen [0]{}%
\providecommand \bibitemStop [0]{}%
\providecommand \bibitemNoStop [0]{.\EOS\space}%
\providecommand \EOS [0]{\spacefactor3000\relax}%
\providecommand \BibitemShut  [1]{\csname bibitem#1\endcsname}%
\let\auto@bib@innerbib\@empty
%</preamble>
\bibitem [{\citenamefont {Bertone}\ \emph {et~al.}(2005)\citenamefont
  {Bertone}, \citenamefont {Hooper},\ and\ \citenamefont
  {Silk}}]{Bertone_2005_Review}%
  \BibitemOpen
  \bibfield  {author} {\bibinfo {author} {\bibfnamefont {G.}~\bibnamefont
  {Bertone}}, \bibinfo {author} {\bibfnamefont {D.}~\bibnamefont {Hooper}},\
  and\ \bibinfo {author} {\bibfnamefont {J.}~\bibnamefont {Silk}},\ }\bibfield
  {title} {\bibinfo {title} {Particle dark matter: evidence, candidates and
  constraints},\ }\href
  {https://doi.org/https://doi.org/10.1016/j.physrep.2004.08.031} {\bibfield
  {journal} {\bibinfo  {journal} {Physics Reports}\ }\textbf {\bibinfo {volume}
  {405}},\ \bibinfo {pages} {279} (\bibinfo {year} {2005})}\BibitemShut
  {NoStop}%
\bibitem [{\citenamefont {Bertone}\ and\ \citenamefont
  {Hooper}(2018)}]{bertone_history_2018-1}%
  \BibitemOpen
  \bibfield  {author} {\bibinfo {author} {\bibfnamefont {G.}~\bibnamefont
  {Bertone}}\ and\ \bibinfo {author} {\bibfnamefont {D.}~\bibnamefont
  {Hooper}},\ }\bibfield  {title} {\bibinfo {title} {History of dark matter},\
  }\href {https://doi.org/10.1103/RevModPhys.90.045002} {\bibfield  {journal}
  {\bibinfo  {journal} {Rev. Mod. Phys.}\ }\textbf {\bibinfo {volume} {90}},\
  \bibinfo {pages} {045002} (\bibinfo {year} {2018})}\BibitemShut {NoStop}%
\bibitem [{\citenamefont {Alvey}\ and\ \citenamefont
  {Fairbairn}(2019)}]{Alvey_2019}%
  \BibitemOpen
  \bibfield  {author} {\bibinfo {author} {\bibfnamefont {J.}~\bibnamefont
  {Alvey}}\ and\ \bibinfo {author} {\bibfnamefont {M.}~\bibnamefont
  {Fairbairn}},\ }\bibfield  {title} {\bibinfo {title} {Linking scalar dark
  matter and neutrino masses with icecube 170922a},\ }\href
  {https://doi.org/10.1088/1475-7516/2019/07/041} {\bibfield  {journal}
  {\bibinfo  {journal} {Journal of Cosmology and Astroparticle Physics}\
  }\textbf {\bibinfo {volume} {2019}}\bibinfo  {number} { (07)},\ \bibinfo
  {pages} {041}}\BibitemShut {NoStop}%
\bibitem [{\citenamefont {Safronova}\ \emph {et~al.}(2018)\citenamefont
  {Safronova}, \citenamefont {Budker}, \citenamefont {DeMille}, \citenamefont
  {Kimball}, \citenamefont {Derevianko},\ and\ \citenamefont
  {Clark}}]{Safronova2018}%
  \BibitemOpen
\bibfield  {number} {  }\bibfield  {author} {\bibinfo {author} {\bibfnamefont
  {M.~S.}\ \bibnamefont {Safronova}}, \bibinfo {author} {\bibfnamefont
  {D.}~\bibnamefont {Budker}}, \bibinfo {author} {\bibfnamefont
  {D.}~\bibnamefont {DeMille}}, \bibinfo {author} {\bibfnamefont {D.~F.~J.}\
  \bibnamefont {Kimball}}, \bibinfo {author} {\bibfnamefont {A.}~\bibnamefont
  {Derevianko}},\ and\ \bibinfo {author} {\bibfnamefont {C.~W.}\ \bibnamefont
  {Clark}},\ }\bibfield  {title} {\bibinfo {title} {Search for new physics with
  atoms and molecules},\ }\href {https://doi.org/10.1103/RevModPhys.90.025008}
  {\bibfield  {journal} {\bibinfo  {journal} {Rev. Mod. Phys.}\ }\textbf
  {\bibinfo {volume} {90}},\ \bibinfo {pages} {025008} (\bibinfo {year}
  {2018})}\BibitemShut {NoStop}%
\bibitem [{\citenamefont {Frugiuele}\ \emph {et~al.}(2017)\citenamefont
  {Frugiuele}, \citenamefont {Fuchs}, \citenamefont {Perez},\ and\
  \citenamefont {Schlaffer}}]{Frugiuele2017}%
  \BibitemOpen
  \bibfield  {author} {\bibinfo {author} {\bibfnamefont {C.}~\bibnamefont
  {Frugiuele}}, \bibinfo {author} {\bibfnamefont {E.}~\bibnamefont {Fuchs}},
  \bibinfo {author} {\bibfnamefont {G.}~\bibnamefont {Perez}},\ and\ \bibinfo
  {author} {\bibfnamefont {M.}~\bibnamefont {Schlaffer}},\ }\bibfield  {title}
  {\bibinfo {title} {Constraining new physics models with isotope shift
  spectroscopy},\ }\href {https://doi.org/10.1103/PhysRevD.96.015011}
  {\bibfield  {journal} {\bibinfo  {journal} {Phys. Rev. D}\ }\textbf {\bibinfo
  {volume} {96}},\ \bibinfo {pages} {015011} (\bibinfo {year}
  {2017})}\BibitemShut {NoStop}%
\bibitem [{\citenamefont {Berengut}\ \emph {et~al.}(2018)\citenamefont
  {Berengut}, \citenamefont {Budker}, \citenamefont {Delaunay}, \citenamefont
  {Flambaum}, \citenamefont {Frugiuele}, \citenamefont {Fuchs}, \citenamefont
  {Grojean}, \citenamefont {Harnik}, \citenamefont {Ozeri}, \citenamefont
  {Perez},\ and\ \citenamefont {Soreq}}]{Berengut2018}%
  \BibitemOpen
  \bibfield  {author} {\bibinfo {author} {\bibfnamefont {J.~C.}\ \bibnamefont
  {Berengut}}, \bibinfo {author} {\bibfnamefont {D.}~\bibnamefont {Budker}},
  \bibinfo {author} {\bibfnamefont {C.}~\bibnamefont {Delaunay}}, \bibinfo
  {author} {\bibfnamefont {V.~V.}\ \bibnamefont {Flambaum}}, \bibinfo {author}
  {\bibfnamefont {C.}~\bibnamefont {Frugiuele}}, \bibinfo {author}
  {\bibfnamefont {E.}~\bibnamefont {Fuchs}}, \bibinfo {author} {\bibfnamefont
  {C.}~\bibnamefont {Grojean}}, \bibinfo {author} {\bibfnamefont
  {R.}~\bibnamefont {Harnik}}, \bibinfo {author} {\bibfnamefont
  {R.}~\bibnamefont {Ozeri}}, \bibinfo {author} {\bibfnamefont
  {G.}~\bibnamefont {Perez}},\ and\ \bibinfo {author} {\bibfnamefont
  {Y.}~\bibnamefont {Soreq}},\ }\bibfield  {title} {\bibinfo {title} {Probing
  new long-range interactions by isotope shift spectroscopy},\ }\href
  {https://doi.org/10.1103/PhysRevLett.120.091801} {\bibfield  {journal}
  {\bibinfo  {journal} {Phys. Rev. Lett.}\ }\textbf {\bibinfo {volume} {120}},\
  \bibinfo {pages} {091801} (\bibinfo {year} {2018})}\BibitemShut {NoStop}%
\bibitem [{\citenamefont {King}(1963)}]{king_1963}%
  \BibitemOpen
  \bibfield  {author} {\bibinfo {author} {\bibfnamefont {W.~H.}\ \bibnamefont
  {King}},\ }\bibfield  {title} {\bibinfo {title} {Comments on the article
  ``peculiarities of the isotope shift in the samarium spectrum''},\ }\href
  {https://doi.org/10.1364/JOSA.53.000638} {\bibfield  {journal} {\bibinfo
  {journal} {J. Opt. Soc. Am.}\ }\textbf {\bibinfo {volume} {53}},\ \bibinfo
  {pages} {638} (\bibinfo {year} {1963})}\BibitemShut {NoStop}%
\bibitem [{\citenamefont {Flambaum}\ \emph {et~al.}(2018)\citenamefont
  {Flambaum}, \citenamefont {Geddes},\ and\ \citenamefont
  {Viatkina}}]{Flambaum2018}%
  \BibitemOpen
  \bibfield  {author} {\bibinfo {author} {\bibfnamefont {V.~V.}\ \bibnamefont
  {Flambaum}}, \bibinfo {author} {\bibfnamefont {A.~J.}\ \bibnamefont
  {Geddes}},\ and\ \bibinfo {author} {\bibfnamefont {A.~V.}\ \bibnamefont
  {Viatkina}},\ }\bibfield  {title} {\bibinfo {title} {Isotope shift,
  nonlinearity of king plots, and the search for new particles},\ }\href
  {https://doi.org/10.1103/PhysRevA.97.032510} {\bibfield  {journal} {\bibinfo
  {journal} {Phys. Rev. A}\ }\textbf {\bibinfo {volume} {97}},\ \bibinfo
  {pages} {032510} (\bibinfo {year} {2018})}\BibitemShut {NoStop}%
\bibitem [{\citenamefont {Fichet}(2018)}]{Fichet2018}%
  \BibitemOpen
  \bibfield  {author} {\bibinfo {author} {\bibfnamefont {S.}~\bibnamefont
  {Fichet}},\ }\bibfield  {title} {\bibinfo {title} {Quantum forces from dark
  matter and where to find them},\ }\href
  {https://doi.org/10.1103/PhysRevLett.120.131801} {\bibfield  {journal}
  {\bibinfo  {journal} {Phys. Rev. Lett.}\ }\textbf {\bibinfo {volume} {120}},\
  \bibinfo {pages} {131801} (\bibinfo {year} {2018})}\BibitemShut {NoStop}%
\bibitem [{\citenamefont {Drake}\ \emph {et~al.}(2021)\citenamefont {Drake},
  \citenamefont {Dhindsa},\ and\ \citenamefont {Marton}}]{Drake2021}%
  \BibitemOpen
  \bibfield  {author} {\bibinfo {author} {\bibfnamefont {G.~W.~F.}\
  \bibnamefont {Drake}}, \bibinfo {author} {\bibfnamefont {H.~S.}\ \bibnamefont
  {Dhindsa}},\ and\ \bibinfo {author} {\bibfnamefont {V.~J.}\ \bibnamefont
  {Marton}},\ }\bibfield  {title} {\bibinfo {title} {King and second-king plots
  with optimized sensitivity for lithium ions},\ }\href
  {https://doi.org/10.1103/PhysRevA.104.L060801} {\bibfield  {journal}
  {\bibinfo  {journal} {Phys. Rev. A}\ }\textbf {\bibinfo {volume} {104}},\
  \bibinfo {pages} {L060801} (\bibinfo {year} {2021})}\BibitemShut {NoStop}%
\bibitem [{\citenamefont {Dhindsa}\ \emph {et~al.}(2022)\citenamefont
  {Dhindsa}, \citenamefont {Marton},\ and\ \citenamefont
  {Drake}}]{Dhindsa2022}%
  \BibitemOpen
  \bibfield  {author} {\bibinfo {author} {\bibfnamefont {H.~S.}\ \bibnamefont
  {Dhindsa}}, \bibinfo {author} {\bibfnamefont {V.~J.}\ \bibnamefont
  {Marton}},\ and\ \bibinfo {author} {\bibfnamefont {G.~W.~F.}\ \bibnamefont
  {Drake}},\ }\bibfield  {title} {\bibinfo {title} {Search for light bosons
  with king and second-king plots optimized for lithium ions},\ }\href
  {https://doi.org/10.1134/S1063779622040037} {\bibfield  {journal} {\bibinfo
  {journal} {Physics of Particles and Nuclei}\ }\textbf {\bibinfo {volume}
  {53}},\ \bibinfo {pages} {800} (\bibinfo {year} {2022})}\BibitemShut
  {NoStop}%
\bibitem [{\citenamefont {Mikami}\ \emph {et~al.}(2017)\citenamefont {Mikami},
  \citenamefont {Tanaka},\ and\ \citenamefont {Yamamoto}}]{Mikami2017}%
  \BibitemOpen
  \bibfield  {author} {\bibinfo {author} {\bibfnamefont {K.}~\bibnamefont
  {Mikami}}, \bibinfo {author} {\bibfnamefont {M.}~\bibnamefont {Tanaka}},\
  and\ \bibinfo {author} {\bibfnamefont {Y.}~\bibnamefont {Yamamoto}},\
  }\bibfield  {title} {\bibinfo {title} {Probing new intra-atomic force with
  isotope shifts},\ }\href {https://doi.org/10.1140/epjc/s10052-017-5467-4}
  {\bibfield  {journal} {\bibinfo  {journal} {The European Physical Journal C}\
  }\textbf {\bibinfo {volume} {77}},\ \bibinfo {pages} {896} (\bibinfo {year}
  {2017})}\BibitemShut {NoStop}%
\bibitem [{\citenamefont {Berengut}\ \emph {et~al.}(2020)\citenamefont
  {Berengut}, \citenamefont {Delaunay}, \citenamefont {Geddes},\ and\
  \citenamefont {Soreq}}]{berengut_2020}%
  \BibitemOpen
  \bibfield  {author} {\bibinfo {author} {\bibfnamefont {J.~C.}\ \bibnamefont
  {Berengut}}, \bibinfo {author} {\bibfnamefont {C.}~\bibnamefont {Delaunay}},
  \bibinfo {author} {\bibfnamefont {A.}~\bibnamefont {Geddes}},\ and\ \bibinfo
  {author} {\bibfnamefont {Y.}~\bibnamefont {Soreq}},\ }\bibfield  {title}
  {\bibinfo {title} {Generalized king linearity and new physics searches with
  isotope shifts},\ }\href {https://doi.org/10.1103/PhysRevResearch.2.043444}
  {\bibfield  {journal} {\bibinfo  {journal} {Phys. Rev. Research}\ }\textbf
  {\bibinfo {volume} {2}},\ \bibinfo {pages} {043444} (\bibinfo {year}
  {2020})}\BibitemShut {NoStop}%
\bibitem [{\citenamefont {Counts}\ \emph {et~al.}(2020)\citenamefont {Counts},
  \citenamefont {Hur}, \citenamefont {Aude~Craik}, \citenamefont {Jeon},
  \citenamefont {Leung}, \citenamefont {Berengut}, \citenamefont {Geddes},
  \citenamefont {Kawasaki}, \citenamefont {Jhe},\ and\ \citenamefont
  {Vuleti\ifmmode~\acute{c}\else \'{c}\fi{}}}]{counts_observation_2020}%
  \BibitemOpen
  \bibfield  {author} {\bibinfo {author} {\bibfnamefont {I.}~\bibnamefont
  {Counts}}, \bibinfo {author} {\bibfnamefont {J.}~\bibnamefont {Hur}},
  \bibinfo {author} {\bibfnamefont {D.~P.~L.}\ \bibnamefont {Aude~Craik}},
  \bibinfo {author} {\bibfnamefont {H.}~\bibnamefont {Jeon}}, \bibinfo {author}
  {\bibfnamefont {C.}~\bibnamefont {Leung}}, \bibinfo {author} {\bibfnamefont
  {J.~C.}\ \bibnamefont {Berengut}}, \bibinfo {author} {\bibfnamefont
  {A.}~\bibnamefont {Geddes}}, \bibinfo {author} {\bibfnamefont
  {A.}~\bibnamefont {Kawasaki}}, \bibinfo {author} {\bibfnamefont
  {W.}~\bibnamefont {Jhe}},\ and\ \bibinfo {author} {\bibfnamefont
  {V.}~\bibnamefont {Vuleti\ifmmode~\acute{c}\else \'{c}\fi{}}},\ }\bibfield
  {title} {\bibinfo {title} {Evidence for nonlinear isotope shift in
  {${\mathrm{Yb}}^{+}$} search for new boson},\ }\href
  {https://doi.org/10.1103/PhysRevLett.125.123002} {\bibfield  {journal}
  {\bibinfo  {journal} {Phys. Rev. Lett.}\ }\textbf {\bibinfo {volume} {125}},\
  \bibinfo {pages} {123002} (\bibinfo {year} {2020})}\BibitemShut {NoStop}%
\bibitem [{\citenamefont {Figueroa}\ \emph {et~al.}(2022)\citenamefont
  {Figueroa}, \citenamefont {Berengut}, \citenamefont {Dzuba}, \citenamefont
  {Flambaum}, \citenamefont {Budker},\ and\ \citenamefont
  {Antypas}}]{Figueroa_2022}%
  \BibitemOpen
  \bibfield  {author} {\bibinfo {author} {\bibfnamefont {N.~L.}\ \bibnamefont
  {Figueroa}}, \bibinfo {author} {\bibfnamefont {J.~C.}\ \bibnamefont
  {Berengut}}, \bibinfo {author} {\bibfnamefont {V.~A.}\ \bibnamefont {Dzuba}},
  \bibinfo {author} {\bibfnamefont {V.~V.}\ \bibnamefont {Flambaum}}, \bibinfo
  {author} {\bibfnamefont {D.}~\bibnamefont {Budker}},\ and\ \bibinfo {author}
  {\bibfnamefont {D.}~\bibnamefont {Antypas}},\ }\bibfield  {title} {\bibinfo
  {title} {Precision determination of isotope shifts in ytterbium and
  implications for new physics},\ }\href
  {https://doi.org/10.1103/PhysRevLett.128.073001} {\bibfield  {journal}
  {\bibinfo  {journal} {Phys. Rev. Lett.}\ }\textbf {\bibinfo {volume} {128}},\
  \bibinfo {pages} {073001} (\bibinfo {year} {2022})}\BibitemShut {NoStop}%
\bibitem [{\citenamefont {Ono}\ \emph {et~al.}(2022)\citenamefont {Ono},
  \citenamefont {Saito}, \citenamefont {Ishiyama}, \citenamefont {Higomoto},
  \citenamefont {Takano}, \citenamefont {Takasu}, \citenamefont {Yamamoto},
  \citenamefont {Tanaka},\ and\ \citenamefont {Takahashi}}]{ono22prx}%
  \BibitemOpen
  \bibfield  {author} {\bibinfo {author} {\bibfnamefont {K.}~\bibnamefont
  {Ono}}, \bibinfo {author} {\bibfnamefont {Y.}~\bibnamefont {Saito}}, \bibinfo
  {author} {\bibfnamefont {T.}~\bibnamefont {Ishiyama}}, \bibinfo {author}
  {\bibfnamefont {T.}~\bibnamefont {Higomoto}}, \bibinfo {author}
  {\bibfnamefont {T.}~\bibnamefont {Takano}}, \bibinfo {author} {\bibfnamefont
  {Y.}~\bibnamefont {Takasu}}, \bibinfo {author} {\bibfnamefont
  {Y.}~\bibnamefont {Yamamoto}}, \bibinfo {author} {\bibfnamefont
  {M.}~\bibnamefont {Tanaka}},\ and\ \bibinfo {author} {\bibfnamefont
  {Y.}~\bibnamefont {Takahashi}},\ }\bibfield  {title} {\bibinfo {title}
  {Observation of nonlinearity of generalized king plot in the search for new
  boson},\ }\href {https://doi.org/10.1103/PhysRevX.12.021033} {\bibfield
  {journal} {\bibinfo  {journal} {Phys. Rev. X}\ }\textbf {\bibinfo {volume}
  {12}},\ \bibinfo {pages} {021033} (\bibinfo {year} {2022})}\BibitemShut
  {NoStop}%
\bibitem [{\citenamefont {Hur}\ \emph {et~al.}(2022)\citenamefont {Hur},
  \citenamefont {Aude~Craik}, \citenamefont {Counts}, \citenamefont {Knyazev},
  \citenamefont {Caldwell}, \citenamefont {Leung}, \citenamefont {Pandey},
  \citenamefont {Berengut}, \citenamefont {Geddes}, \citenamefont {Nazarewicz},
  \citenamefont {Reinhard}, \citenamefont {Kawasaki}, \citenamefont {Jeon},
  \citenamefont {Jhe},\ and\ \citenamefont {Vuleti\ifmmode~\acute{c}\else
  \'{c}\fi{}}}]{hur22prl}%
  \BibitemOpen
  \bibfield  {author} {\bibinfo {author} {\bibfnamefont {J.}~\bibnamefont
  {Hur}}, \bibinfo {author} {\bibfnamefont {D.~P.~L.}\ \bibnamefont
  {Aude~Craik}}, \bibinfo {author} {\bibfnamefont {I.}~\bibnamefont {Counts}},
  \bibinfo {author} {\bibfnamefont {E.}~\bibnamefont {Knyazev}}, \bibinfo
  {author} {\bibfnamefont {L.}~\bibnamefont {Caldwell}}, \bibinfo {author}
  {\bibfnamefont {C.}~\bibnamefont {Leung}}, \bibinfo {author} {\bibfnamefont
  {S.}~\bibnamefont {Pandey}}, \bibinfo {author} {\bibfnamefont {J.~C.}\
  \bibnamefont {Berengut}}, \bibinfo {author} {\bibfnamefont {A.}~\bibnamefont
  {Geddes}}, \bibinfo {author} {\bibfnamefont {W.}~\bibnamefont {Nazarewicz}},
  \bibinfo {author} {\bibfnamefont {P.-G.}\ \bibnamefont {Reinhard}}, \bibinfo
  {author} {\bibfnamefont {A.}~\bibnamefont {Kawasaki}}, \bibinfo {author}
  {\bibfnamefont {H.}~\bibnamefont {Jeon}}, \bibinfo {author} {\bibfnamefont
  {W.}~\bibnamefont {Jhe}},\ and\ \bibinfo {author} {\bibfnamefont
  {V.}~\bibnamefont {Vuleti\ifmmode~\acute{c}\else \'{c}\fi{}}},\ }\bibfield
  {title} {\bibinfo {title} {Evidence of two-source king plot nonlinearity in
  spectroscopic search for new boson},\ }\href
  {https://doi.org/10.1103/PhysRevLett.128.163201} {\bibfield  {journal}
  {\bibinfo  {journal} {Phys. Rev. Lett.}\ }\textbf {\bibinfo {volume} {128}},\
  \bibinfo {pages} {163201} (\bibinfo {year} {2022})}\BibitemShut {NoStop}%
\bibitem [{\citenamefont {Allehabi}\ \emph {et~al.}(2021)\citenamefont
  {Allehabi}, \citenamefont {Dzuba}, \citenamefont {Flambaum},\ and\
  \citenamefont {Afanasjev}}]{Allehabi_2021}%
  \BibitemOpen
  \bibfield  {author} {\bibinfo {author} {\bibfnamefont {S.~O.}\ \bibnamefont
  {Allehabi}}, \bibinfo {author} {\bibfnamefont {V.~A.}\ \bibnamefont {Dzuba}},
  \bibinfo {author} {\bibfnamefont {V.~V.}\ \bibnamefont {Flambaum}},\ and\
  \bibinfo {author} {\bibfnamefont {A.~V.}\ \bibnamefont {Afanasjev}},\
  }\bibfield  {title} {\bibinfo {title} {Nuclear deformation as a source of the
  nonlinearity of the king plot in the {${\mathrm{Yb}}^{+}$} ion},\ }\href
  {https://doi.org/10.1103/PhysRevA.103.L030801} {\bibfield  {journal}
  {\bibinfo  {journal} {Phys. Rev. A}\ }\textbf {\bibinfo {volume} {103}},\
  \bibinfo {pages} {L030801} (\bibinfo {year} {2021})}\BibitemShut {NoStop}%
\bibitem [{\citenamefont {Solaro}\ \emph {et~al.}(2020)\citenamefont {Solaro},
  \citenamefont {Meyer}, \citenamefont {Fisher}, \citenamefont {Berengut},
  \citenamefont {Fuchs},\ and\ \citenamefont {Drewsen}}]{Solaro2020}%
  \BibitemOpen
  \bibfield  {author} {\bibinfo {author} {\bibfnamefont {C.}~\bibnamefont
  {Solaro}}, \bibinfo {author} {\bibfnamefont {S.}~\bibnamefont {Meyer}},
  \bibinfo {author} {\bibfnamefont {K.}~\bibnamefont {Fisher}}, \bibinfo
  {author} {\bibfnamefont {J.~C.}\ \bibnamefont {Berengut}}, \bibinfo {author}
  {\bibfnamefont {E.}~\bibnamefont {Fuchs}},\ and\ \bibinfo {author}
  {\bibfnamefont {M.}~\bibnamefont {Drewsen}},\ }\bibfield  {title} {\bibinfo
  {title} {Improved isotope-shift-based bounds on bosons beyond the standard
  model through measurements of the
  ${^{2}\mathrm{D}}_{3/2}\ensuremath{-}{^{2}\mathrm{D}}_{5/2}$ interval in
  {${\mathrm{Ca}}^{+}$}},\ }\href
  {https://doi.org/10.1103/PhysRevLett.125.123003} {\bibfield  {journal}
  {\bibinfo  {journal} {Phys. Rev. Lett.}\ }\textbf {\bibinfo {volume} {125}},\
  \bibinfo {pages} {123003} (\bibinfo {year} {2020})}\BibitemShut {NoStop}%
\bibitem [{\citenamefont {Shabaev}\ \emph {et~al.}(2018)\citenamefont
  {Shabaev}, \citenamefont {Bondarev}, \citenamefont {Glazov}, \citenamefont
  {Kaygorodov}, \citenamefont {Kozhedub}, \citenamefont {Maltsev},
  \citenamefont {Malyshev}, \citenamefont {Popov}, \citenamefont {Tupitsyn},\
  and\ \citenamefont {Zubova}}]{Shabaev2018}%
  \BibitemOpen
  \bibfield  {author} {\bibinfo {author} {\bibfnamefont {V.~M.}\ \bibnamefont
  {Shabaev}}, \bibinfo {author} {\bibfnamefont {A.~I.}\ \bibnamefont
  {Bondarev}}, \bibinfo {author} {\bibfnamefont {D.~A.}\ \bibnamefont
  {Glazov}}, \bibinfo {author} {\bibfnamefont {M.~Y.}\ \bibnamefont
  {Kaygorodov}}, \bibinfo {author} {\bibfnamefont {Y.~S.}\ \bibnamefont
  {Kozhedub}}, \bibinfo {author} {\bibfnamefont {I.~A.}\ \bibnamefont
  {Maltsev}}, \bibinfo {author} {\bibfnamefont {A.~V.}\ \bibnamefont
  {Malyshev}}, \bibinfo {author} {\bibfnamefont {R.~V.}\ \bibnamefont {Popov}},
  \bibinfo {author} {\bibfnamefont {I.~I.}\ \bibnamefont {Tupitsyn}},\ and\
  \bibinfo {author} {\bibfnamefont {N.~A.}\ \bibnamefont {Zubova}},\ }\bibfield
   {title} {\bibinfo {title} {Stringent tests of {QED} using highly charged
  ions},\ }\href {https://doi.org/10.1007/s10751-018-1537-8} {\bibfield
  {journal} {\bibinfo  {journal} {Hyperfine Interactions}\ }\textbf {\bibinfo
  {volume} {239}},\ \bibinfo {pages} {60} (\bibinfo {year} {2018})}\BibitemShut
  {NoStop}%
\bibitem [{\citenamefont {Indelicato}(2019)}]{Indelicato_2019}%
  \BibitemOpen
  \bibfield  {author} {\bibinfo {author} {\bibfnamefont {P.}~\bibnamefont
  {Indelicato}},\ }\bibfield  {title} {\bibinfo {title} {{QED} tests with
  highly charged ions},\ }\href {https://doi.org/10.1088/1361-6455/ab42c9}
  {\bibfield  {journal} {\bibinfo  {journal} {Journal of Physics B: Atomic,
  Molecular and Optical Physics}\ }\textbf {\bibinfo {volume} {52}},\ \bibinfo
  {pages} {232001} (\bibinfo {year} {2019})}\BibitemShut {NoStop}%
\bibitem [{\citenamefont {Kozlov}\ \emph {et~al.}(2018)\citenamefont {Kozlov},
  \citenamefont {Safronova}, \citenamefont {Crespo L\'opez-Urrutia},\ and\
  \citenamefont {Schmidt}}]{KozlovRMP2018}%
  \BibitemOpen
  \bibfield  {author} {\bibinfo {author} {\bibfnamefont {M.~G.}\ \bibnamefont
  {Kozlov}}, \bibinfo {author} {\bibfnamefont {M.~S.}\ \bibnamefont
  {Safronova}}, \bibinfo {author} {\bibfnamefont {J.~R.}\ \bibnamefont {Crespo
  L\'opez-Urrutia}},\ and\ \bibinfo {author} {\bibfnamefont {P.~O.}\
  \bibnamefont {Schmidt}},\ }\bibfield  {title} {\bibinfo {title} {Highly
  charged ions: Optical clocks and applications in fundamental physics},\
  }\href {https://doi.org/10.1103/RevModPhys.90.045005} {\bibfield  {journal}
  {\bibinfo  {journal} {Rev. Mod. Phys.}\ }\textbf {\bibinfo {volume} {90}},\
  \bibinfo {pages} {045005} (\bibinfo {year} {2018})}\BibitemShut {NoStop}%
\bibitem [{\citenamefont {Schmöger}\ \emph {et~al.}(2015)\citenamefont
  {Schmöger}, \citenamefont {Versolato}, \citenamefont {Schwarz},
  \citenamefont {Kohnen}, \citenamefont {Windberger}, \citenamefont {Piest},
  \citenamefont {Feuchtenbeiner}, \citenamefont {Pedregosa-Gutierrez},
  \citenamefont {Leopold}, \citenamefont {Micke}, \citenamefont {Hansen},
  \citenamefont {Baumann}, \citenamefont {Drewsen}, \citenamefont {Ullrich},
  \citenamefont {Schmidt},\ and\ \citenamefont
  {López-Urrutia}}]{schmoger_coulomb_2015}%
  \BibitemOpen
  \bibfield  {author} {\bibinfo {author} {\bibfnamefont {L.}~\bibnamefont
  {Schmöger}}, \bibinfo {author} {\bibfnamefont {O.~O.}\ \bibnamefont
  {Versolato}}, \bibinfo {author} {\bibfnamefont {M.}~\bibnamefont {Schwarz}},
  \bibinfo {author} {\bibfnamefont {M.}~\bibnamefont {Kohnen}}, \bibinfo
  {author} {\bibfnamefont {A.}~\bibnamefont {Windberger}}, \bibinfo {author}
  {\bibfnamefont {B.}~\bibnamefont {Piest}}, \bibinfo {author} {\bibfnamefont
  {S.}~\bibnamefont {Feuchtenbeiner}}, \bibinfo {author} {\bibfnamefont
  {J.}~\bibnamefont {Pedregosa-Gutierrez}}, \bibinfo {author} {\bibfnamefont
  {T.}~\bibnamefont {Leopold}}, \bibinfo {author} {\bibfnamefont
  {P.}~\bibnamefont {Micke}}, \bibinfo {author} {\bibfnamefont {A.~K.}\
  \bibnamefont {Hansen}}, \bibinfo {author} {\bibfnamefont {T.~M.}\
  \bibnamefont {Baumann}}, \bibinfo {author} {\bibfnamefont {M.}~\bibnamefont
  {Drewsen}}, \bibinfo {author} {\bibfnamefont {J.}~\bibnamefont {Ullrich}},
  \bibinfo {author} {\bibfnamefont {P.~O.}\ \bibnamefont {Schmidt}},\ and\
  \bibinfo {author} {\bibfnamefont {J.~R.~C.}\ \bibnamefont {López-Urrutia}},\
  }\bibfield  {title} {\bibinfo {title} {Coulomb crystallization of highly
  charged ions},\ }\href {https://doi.org/10.1126/science.aaa2960} {\bibfield
  {journal} {\bibinfo  {journal} {Science}\ }\textbf {\bibinfo {volume}
  {347}},\ \bibinfo {pages} {1233} (\bibinfo {year} {2015})}\BibitemShut
  {NoStop}%
\bibitem [{\citenamefont {Schmidt}\ \emph {et~al.}(2005)\citenamefont
  {Schmidt}, \citenamefont {Rosenband}, \citenamefont {Langer}, \citenamefont
  {Itano}, \citenamefont {Bergquist},\ and\ \citenamefont
  {Wineland}}]{schmidt_2005}%
  \BibitemOpen
  \bibfield  {author} {\bibinfo {author} {\bibfnamefont {P.~O.}\ \bibnamefont
  {Schmidt}}, \bibinfo {author} {\bibfnamefont {T.}~\bibnamefont {Rosenband}},
  \bibinfo {author} {\bibfnamefont {C.}~\bibnamefont {Langer}}, \bibinfo
  {author} {\bibfnamefont {W.~M.}\ \bibnamefont {Itano}}, \bibinfo {author}
  {\bibfnamefont {J.~C.}\ \bibnamefont {Bergquist}},\ and\ \bibinfo {author}
  {\bibfnamefont {D.~J.}\ \bibnamefont {Wineland}},\ }\bibfield  {title}
  {\bibinfo {title} {Spectroscopy using quantum logic},\ }\href
  {https://doi.org/10.1126/science.1114375} {\bibfield  {journal} {\bibinfo
  {journal} {Science}\ }\textbf {\bibinfo {volume} {309}},\ \bibinfo {pages}
  {749} (\bibinfo {year} {2005})}\BibitemShut {NoStop}%
\bibitem [{\citenamefont {Micke}\ \emph {et~al.}(2020)\citenamefont {Micke},
  \citenamefont {Leopold}, \citenamefont {King}, \citenamefont {Benkler},
  \citenamefont {Spie{\ss}}, \citenamefont {Schm{\"o}ger}, \citenamefont
  {Schwarz}, \citenamefont {{Crespo L{\'o}pez-Urrutia}},\ and\ \citenamefont
  {Schmidt}}]{micke_2020}%
  \BibitemOpen
  \bibfield  {author} {\bibinfo {author} {\bibfnamefont {P.}~\bibnamefont
  {Micke}}, \bibinfo {author} {\bibfnamefont {T.}~\bibnamefont {Leopold}},
  \bibinfo {author} {\bibfnamefont {S.~A.}\ \bibnamefont {King}}, \bibinfo
  {author} {\bibfnamefont {E.}~\bibnamefont {Benkler}}, \bibinfo {author}
  {\bibfnamefont {L.~J.}\ \bibnamefont {Spie{\ss}}}, \bibinfo {author}
  {\bibfnamefont {L.}~\bibnamefont {Schm{\"o}ger}}, \bibinfo {author}
  {\bibfnamefont {M.}~\bibnamefont {Schwarz}}, \bibinfo {author} {\bibfnamefont
  {J.~R.}\ \bibnamefont {{Crespo L{\'o}pez-Urrutia}}},\ and\ \bibinfo {author}
  {\bibfnamefont {P.~O.}\ \bibnamefont {Schmidt}},\ }\bibfield  {title}
  {\bibinfo {title} {Coherent laser spectroscopy of highly charged ions using
  quantum logic},\ }\href {https://doi.org/10.1038/s41586-020-1959-8}
  {\bibfield  {journal} {\bibinfo  {journal} {Nature}\ }\textbf {\bibinfo
  {volume} {578}},\ \bibinfo {pages} {60} (\bibinfo {year} {2020})}\BibitemShut
  {NoStop}%
\bibitem [{\citenamefont {King}\ \emph {et~al.}(2021)\citenamefont {King},
  \citenamefont {Spie\ss{}}, \citenamefont {Micke}, \citenamefont {Wilzewski},
  \citenamefont {Leopold}, \citenamefont {Crespo L\'opez-Urrutia},\ and\
  \citenamefont {Schmidt}}]{king_algorithmic_2021}%
  \BibitemOpen
  \bibfield  {author} {\bibinfo {author} {\bibfnamefont {S.~A.}\ \bibnamefont
  {King}}, \bibinfo {author} {\bibfnamefont {L.~J.}\ \bibnamefont {Spie\ss{}}},
  \bibinfo {author} {\bibfnamefont {P.}~\bibnamefont {Micke}}, \bibinfo
  {author} {\bibfnamefont {A.}~\bibnamefont {Wilzewski}}, \bibinfo {author}
  {\bibfnamefont {T.}~\bibnamefont {Leopold}}, \bibinfo {author} {\bibfnamefont
  {J.~R.}\ \bibnamefont {Crespo L\'opez-Urrutia}},\ and\ \bibinfo {author}
  {\bibfnamefont {P.~O.}\ \bibnamefont {Schmidt}},\ }\bibfield  {title}
  {\bibinfo {title} {Algorithmic {Ground}-{State} {Cooling} of {Weakly}
  {Coupled} {Oscillators} {Using} {Quantum} {Logic}},\ }\href
  {https://doi.org/10.1103/PhysRevX.11.041049} {\bibfield  {journal} {\bibinfo
  {journal} {Phys. Rev. X}\ }\textbf {\bibinfo {volume} {11}},\ \bibinfo
  {pages} {041049} (\bibinfo {year} {2021})}\BibitemShut {NoStop}%
\bibitem [{\citenamefont {King}\ \emph {et~al.}(2022)\citenamefont {King},
  \citenamefont {Spie{\ss}}, \citenamefont {Micke}, \citenamefont {Wilzewski},
  \citenamefont {Leopold}, \citenamefont {Benkler}, \citenamefont {Lange},
  \citenamefont {Huntemann}, \citenamefont {Surzhykov}, \citenamefont
  {Yerokhin}, \citenamefont {Crespo L{\'o}pez-Urrutia},\ and\ \citenamefont
  {Schmidt}}]{King2022}%
  \BibitemOpen
  \bibfield  {author} {\bibinfo {author} {\bibfnamefont {S.~A.}\ \bibnamefont
  {King}}, \bibinfo {author} {\bibfnamefont {L.~J.}\ \bibnamefont {Spie{\ss}}},
  \bibinfo {author} {\bibfnamefont {P.}~\bibnamefont {Micke}}, \bibinfo
  {author} {\bibfnamefont {A.}~\bibnamefont {Wilzewski}}, \bibinfo {author}
  {\bibfnamefont {T.}~\bibnamefont {Leopold}}, \bibinfo {author} {\bibfnamefont
  {E.}~\bibnamefont {Benkler}}, \bibinfo {author} {\bibfnamefont
  {R.}~\bibnamefont {Lange}}, \bibinfo {author} {\bibfnamefont
  {N.}~\bibnamefont {Huntemann}}, \bibinfo {author} {\bibfnamefont
  {A.}~\bibnamefont {Surzhykov}}, \bibinfo {author} {\bibfnamefont {V.~A.}\
  \bibnamefont {Yerokhin}}, \bibinfo {author} {\bibfnamefont {J.~R.}\
  \bibnamefont {Crespo L{\'o}pez-Urrutia}},\ and\ \bibinfo {author}
  {\bibfnamefont {P.~O.}\ \bibnamefont {Schmidt}},\ }\bibfield  {title}
  {\bibinfo {title} {An optical atomic clock based on a highly charged ion},\
  }\href {https://doi.org/10.1038/s41586-022-05245-4} {\bibfield  {journal}
  {\bibinfo  {journal} {Nature}\ }\textbf {\bibinfo {volume} {611}},\ \bibinfo
  {pages} {43} (\bibinfo {year} {2022})}\BibitemShut {NoStop}%
\bibitem [{\citenamefont {Barbieri}\ and\ \citenamefont
  {Ericson}(1975)}]{barbieri_1975}%
  \BibitemOpen
  \bibfield  {author} {\bibinfo {author} {\bibfnamefont {R.}~\bibnamefont
  {Barbieri}}\ and\ \bibinfo {author} {\bibfnamefont {T.}~\bibnamefont
  {Ericson}},\ }\bibfield  {title} {\bibinfo {title} {Evidence against the
  existence of a low mass scalar boson from neutron-nucleus scattering},\
  }\href {https://doi.org/https://doi.org/10.1016/0370-2693(75)90073-8}
  {\bibfield  {journal} {\bibinfo  {journal} {Physics Letters B}\ }\textbf
  {\bibinfo {volume} {57}},\ \bibinfo {pages} {270 } (\bibinfo {year}
  {1975})}\BibitemShut {NoStop}%
\bibitem [{\citenamefont {Leeb}\ and\ \citenamefont
  {Schmiedmayer}(1992)}]{leeb_1992}%
  \BibitemOpen
  \bibfield  {author} {\bibinfo {author} {\bibfnamefont {H.}~\bibnamefont
  {Leeb}}\ and\ \bibinfo {author} {\bibfnamefont {J.}~\bibnamefont
  {Schmiedmayer}},\ }\bibfield  {title} {\bibinfo {title} {Constraint on
  hypothetical light interacting bosons from low-energy neutron experiments},\
  }\href {https://doi.org/10.1103/PhysRevLett.68.1472} {\bibfield  {journal}
  {\bibinfo  {journal} {Phys. Rev. Lett.}\ }\textbf {\bibinfo {volume} {68}},\
  \bibinfo {pages} {1472} (\bibinfo {year} {1992})}\BibitemShut {NoStop}%
\bibitem [{\citenamefont {Nesvizhevsky}\ \emph {et~al.}(2008)\citenamefont
  {Nesvizhevsky}, \citenamefont {Pignol},\ and\ \citenamefont
  {Protasov}}]{nesvizhevsky_2008}%
  \BibitemOpen
  \bibfield  {author} {\bibinfo {author} {\bibfnamefont {V.~V.}\ \bibnamefont
  {Nesvizhevsky}}, \bibinfo {author} {\bibfnamefont {G.}~\bibnamefont
  {Pignol}},\ and\ \bibinfo {author} {\bibfnamefont {K.~V.}\ \bibnamefont
  {Protasov}},\ }\bibfield  {title} {\bibinfo {title} {Neutron scattering and
  extra-short-range interactions},\ }\href
  {https://doi.org/10.1103/PhysRevD.77.034020} {\bibfield  {journal} {\bibinfo
  {journal} {Phys. Rev. D}\ }\textbf {\bibinfo {volume} {77}},\ \bibinfo
  {pages} {034020} (\bibinfo {year} {2008})}\BibitemShut {NoStop}%
\bibitem [{\citenamefont {Pokotilovski}(2006)}]{pokotilovski_2006}%
  \BibitemOpen
  \bibfield  {author} {\bibinfo {author} {\bibfnamefont {Y.~N.}\ \bibnamefont
  {Pokotilovski}},\ }\bibfield  {title} {\bibinfo {title} {Constraints on new
  interactions from neutron scattering experiments},\ }\href
  {https://doi.org/10.1134/S1063778806060020} {\bibfield  {journal} {\bibinfo
  {journal} {Physics of Atomic Nuclei}\ }\textbf {\bibinfo {volume} {69}},\
  \bibinfo {pages} {924} (\bibinfo {year} {2006})}\BibitemShut {NoStop}%
\bibitem [{\citenamefont {Adler}\ \emph {et~al.}(1974)\citenamefont {Adler},
  \citenamefont {Dashen},\ and\ \citenamefont {Treiman}}]{adler_1974}%
  \BibitemOpen
  \bibfield  {author} {\bibinfo {author} {\bibfnamefont {S.~L.}\ \bibnamefont
  {Adler}}, \bibinfo {author} {\bibfnamefont {R.~F.}\ \bibnamefont {Dashen}},\
  and\ \bibinfo {author} {\bibfnamefont {S.~B.}\ \bibnamefont {Treiman}},\
  }\bibfield  {title} {\bibinfo {title} {Comments on proposed explanations for
  the muonic-atom x-ray discrepancy},\ }\href
  {https://doi.org/10.1103/PhysRevD.10.3728} {\bibfield  {journal} {\bibinfo
  {journal} {Phys. Rev. D}\ }\textbf {\bibinfo {volume} {10}},\ \bibinfo
  {pages} {3728} (\bibinfo {year} {1974})}\BibitemShut {NoStop}%
\bibitem [{\citenamefont {Bordag}\ \emph {et~al.}(2001)\citenamefont {Bordag},
  \citenamefont {Mohideen},\ and\ \citenamefont {Mostepanenko}}]{bordag_2001}%
  \BibitemOpen
  \bibfield  {author} {\bibinfo {author} {\bibfnamefont {M.}~\bibnamefont
  {Bordag}}, \bibinfo {author} {\bibfnamefont {U.}~\bibnamefont {Mohideen}},\
  and\ \bibinfo {author} {\bibfnamefont {V.}~\bibnamefont {Mostepanenko}},\
  }\bibfield  {title} {\bibinfo {title} {New developments in the casimir
  effect},\ }\href
  {https://doi.org/https://doi.org/10.1016/S0370-1573(01)00015-1} {\bibfield
  {journal} {\bibinfo  {journal} {Physics Reports}\ }\textbf {\bibinfo {volume}
  {353}},\ \bibinfo {pages} {1 } (\bibinfo {year} {2001})}\BibitemShut
  {NoStop}%
\bibitem [{\citenamefont {Bordag}\ \emph {et~al.}(2009)\citenamefont {Bordag},
  \citenamefont {Klimchitskaya}, \citenamefont {Mohideen},\ and\ \citenamefont
  {Mostepanenko}}]{bordag_2009}%
  \BibitemOpen
  \bibfield  {author} {\bibinfo {author} {\bibfnamefont {M.}~\bibnamefont
  {Bordag}}, \bibinfo {author} {\bibfnamefont {G.~L.}\ \bibnamefont
  {Klimchitskaya}}, \bibinfo {author} {\bibfnamefont {U.}~\bibnamefont
  {Mohideen}},\ and\ \bibinfo {author} {\bibfnamefont {V.~M.}\ \bibnamefont
  {Mostepanenko}},\ }\href@noop {} {\emph {\bibinfo {title} {Advances in the
  Casimir effect}}},\ Vol.\ \bibinfo {volume} {145}\ (\bibinfo  {publisher}
  {OUP Oxford},\ \bibinfo {year} {2009})\BibitemShut {NoStop}%
\bibitem [{\citenamefont {Germann}\ \emph {et~al.}(2021)\citenamefont
  {Germann}, \citenamefont {Patra}, \citenamefont {Karr}, \citenamefont
  {Hilico}, \citenamefont {Korobov}, \citenamefont {Salumbides}, \citenamefont
  {Eikema}, \citenamefont {Ubachs},\ and\ \citenamefont
  {Koelemeij}}]{germann_three-body_2021}%
  \BibitemOpen
  \bibfield  {author} {\bibinfo {author} {\bibfnamefont {M.}~\bibnamefont
  {Germann}}, \bibinfo {author} {\bibfnamefont {S.}~\bibnamefont {Patra}},
  \bibinfo {author} {\bibfnamefont {J.-P.}\ \bibnamefont {Karr}}, \bibinfo
  {author} {\bibfnamefont {L.}~\bibnamefont {Hilico}}, \bibinfo {author}
  {\bibfnamefont {V.~I.}\ \bibnamefont {Korobov}}, \bibinfo {author}
  {\bibfnamefont {E.~J.}\ \bibnamefont {Salumbides}}, \bibinfo {author}
  {\bibfnamefont {K.~S.~E.}\ \bibnamefont {Eikema}}, \bibinfo {author}
  {\bibfnamefont {W.}~\bibnamefont {Ubachs}},\ and\ \bibinfo {author}
  {\bibfnamefont {J.~C.~J.}\ \bibnamefont {Koelemeij}},\ }\bibfield  {title}
  {\bibinfo {title} {Three-body {{QED}} test and fifth-force constraint from
  vibrations and rotations of \$\{\textbackslash
  mathrm\{\vphantom{\}\}}{{HD}}\vphantom\{\}\vphantom\{\}\^\{+\}\$},\ }\href
  {https://doi.org/10.1103/PhysRevResearch.3.L022028} {\bibfield  {journal}
  {\bibinfo  {journal} {Phys. Rev. Research}\ }\textbf {\bibinfo {volume}
  {3}},\ \bibinfo {pages} {L022028} (\bibinfo {year} {2021})}\BibitemShut
  {NoStop}%
\bibitem [{\citenamefont {Salumbides}\ \emph {et~al.}(2013)\citenamefont
  {Salumbides}, \citenamefont {Koelemeij}, \citenamefont {Komasa},
  \citenamefont {Pachucki}, \citenamefont {Eikema},\ and\ \citenamefont
  {Ubachs}}]{salumbides_bounds_2013}%
  \BibitemOpen
  \bibfield  {author} {\bibinfo {author} {\bibfnamefont {E.~J.}\ \bibnamefont
  {Salumbides}}, \bibinfo {author} {\bibfnamefont {J.~C.~J.}\ \bibnamefont
  {Koelemeij}}, \bibinfo {author} {\bibfnamefont {J.}~\bibnamefont {Komasa}},
  \bibinfo {author} {\bibfnamefont {K.}~\bibnamefont {Pachucki}}, \bibinfo
  {author} {\bibfnamefont {K.~S.~E.}\ \bibnamefont {Eikema}},\ and\ \bibinfo
  {author} {\bibfnamefont {W.}~\bibnamefont {Ubachs}},\ }\bibfield  {title}
  {\bibinfo {title} {Bounds on fifth forces from precision measurements on
  molecules},\ }\href {https://doi.org/10.1103/PhysRevD.87.112008} {\bibfield
  {journal} {\bibinfo  {journal} {Phys. Rev. D}\ }\textbf {\bibinfo {volume}
  {87}},\ \bibinfo {pages} {112008} (\bibinfo {year} {2013})}\BibitemShut
  {NoStop}%
\bibitem [{\citenamefont {Jaffe}\ \emph {et~al.}(2017)\citenamefont {Jaffe},
  \citenamefont {Haslinger}, \citenamefont {Xu}, \citenamefont {Hamilton},
  \citenamefont {Upadhye}, \citenamefont {Elder}, \citenamefont {Khoury},\ and\
  \citenamefont {M{\"u}ller}}]{jaffe_testing_2017}%
  \BibitemOpen
  \bibfield  {author} {\bibinfo {author} {\bibfnamefont {M.}~\bibnamefont
  {Jaffe}}, \bibinfo {author} {\bibfnamefont {P.}~\bibnamefont {Haslinger}},
  \bibinfo {author} {\bibfnamefont {V.}~\bibnamefont {Xu}}, \bibinfo {author}
  {\bibfnamefont {P.}~\bibnamefont {Hamilton}}, \bibinfo {author}
  {\bibfnamefont {A.}~\bibnamefont {Upadhye}}, \bibinfo {author} {\bibfnamefont
  {B.}~\bibnamefont {Elder}}, \bibinfo {author} {\bibfnamefont
  {J.}~\bibnamefont {Khoury}},\ and\ \bibinfo {author} {\bibfnamefont
  {H.}~\bibnamefont {M{\"u}ller}},\ }\bibfield  {title} {\bibinfo {title}
  {Testing sub-gravitational forces on atoms from a miniature in-vacuum source
  mass},\ }\href {https://doi.org/10.1038/nphys4189} {\bibfield  {journal}
  {\bibinfo  {journal} {Nat Phys}\ }\textbf {\bibinfo {volume} {13}},\ \bibinfo
  {pages} {938} (\bibinfo {year} {2017})}\BibitemShut {NoStop}%
\bibitem [{\citenamefont {Biesheuvel}\ \emph {et~al.}(2016)\citenamefont
  {Biesheuvel}, \citenamefont {Karr}, \citenamefont {Hilico}, \citenamefont
  {Eikema}, \citenamefont {Ubachs},\ and\ \citenamefont
  {Koelemeij}}]{biesheuvel_probing_2016}%
  \BibitemOpen
  \bibfield  {author} {\bibinfo {author} {\bibfnamefont {J.}~\bibnamefont
  {Biesheuvel}}, \bibinfo {author} {\bibfnamefont {J.-P.}\ \bibnamefont
  {Karr}}, \bibinfo {author} {\bibfnamefont {L.}~\bibnamefont {Hilico}},
  \bibinfo {author} {\bibfnamefont {K.~S.~E.}\ \bibnamefont {Eikema}}, \bibinfo
  {author} {\bibfnamefont {W.}~\bibnamefont {Ubachs}},\ and\ \bibinfo {author}
  {\bibfnamefont {J.~C.~J.}\ \bibnamefont {Koelemeij}},\ }\bibfield  {title}
  {\bibinfo {title} {Probing {{QED}} and fundamental constants through laser
  spectroscopy of vibrational transitions in {{HD}}{\textsuperscript{+}}},\
  }\href {https://doi.org/10.1038/ncomms10385} {\bibfield  {journal} {\bibinfo
  {journal} {Nat Commun}\ }\textbf {\bibinfo {volume} {7}},\ \bibinfo {pages}
  {10385} (\bibinfo {year} {2016})}\BibitemShut {NoStop}%
\bibitem [{\citenamefont {Kahl}\ and\ \citenamefont
  {Berengut}(2019)}]{ambit_2019}%
  \BibitemOpen
  \bibfield  {author} {\bibinfo {author} {\bibfnamefont {E.}~\bibnamefont
  {Kahl}}\ and\ \bibinfo {author} {\bibfnamefont {J.}~\bibnamefont
  {Berengut}},\ }\bibfield  {title} {\bibinfo {title} {{AMBiT}: A programme for
  high-precision relativistic atomic structure calculations},\ }\href
  {https://doi.org/https://doi.org/10.1016/j.cpc.2018.12.014} {\bibfield
  {journal} {\bibinfo  {journal} {Computer Physics Communications}\ }\textbf
  {\bibinfo {volume} {238}},\ \bibinfo {pages} {232} (\bibinfo {year}
  {2019})}\BibitemShut {NoStop}%
\bibitem [{\citenamefont {Levine}\ \emph {et~al.}(1988)\citenamefont {Levine},
  \citenamefont {Marrs}, \citenamefont {Henderson}, \citenamefont {Knapp},\
  and\ \citenamefont {Schneider}}]{levine_electron_1988}%
  \BibitemOpen
  \bibfield  {author} {\bibinfo {author} {\bibfnamefont {M.~A.}\ \bibnamefont
  {Levine}}, \bibinfo {author} {\bibfnamefont {R.~E.}\ \bibnamefont {Marrs}},
  \bibinfo {author} {\bibfnamefont {J.~R.}\ \bibnamefont {Henderson}}, \bibinfo
  {author} {\bibfnamefont {D.~A.}\ \bibnamefont {Knapp}},\ and\ \bibinfo
  {author} {\bibfnamefont {M.~B.}\ \bibnamefont {Schneider}},\ }\bibfield
  {title} {\bibinfo {title} {The {Electron} {Beam} {Ion} {Trap}: {A} {New}
  {Instrument} for {Atomic} {Physics} {Measurements}},\ }\href
  {https://doi.org/10.1088/0031-8949/1988/T22/024} {\bibfield  {journal}
  {\bibinfo  {journal} {Physica Scripta}\ }\textbf {\bibinfo {volume} {1988}},\
  \bibinfo {pages} {157} (\bibinfo {year} {1988})}\BibitemShut {NoStop}%
\bibitem [{\citenamefont {Levine}\ \emph {et~al.}(1989)\citenamefont {Levine},
  \citenamefont {Marrs}, \citenamefont {Bardsley}, \citenamefont
  {Beiersdorfer}, \citenamefont {Bennett}, \citenamefont {Chen}, \citenamefont
  {Cowan}, \citenamefont {Dietrich}, \citenamefont {Henderson}, \citenamefont
  {Knapp}, \citenamefont {Osterheld}, \citenamefont {Penetrante}, \citenamefont
  {Schneider},\ and\ \citenamefont {Scofield}}]{levine_use_1989}%
  \BibitemOpen
  \bibfield  {author} {\bibinfo {author} {\bibfnamefont {M.~A.}\ \bibnamefont
  {Levine}}, \bibinfo {author} {\bibfnamefont {R.~E.}\ \bibnamefont {Marrs}},
  \bibinfo {author} {\bibfnamefont {J.~N.}\ \bibnamefont {Bardsley}}, \bibinfo
  {author} {\bibfnamefont {P.}~\bibnamefont {Beiersdorfer}}, \bibinfo {author}
  {\bibfnamefont {C.~L.}\ \bibnamefont {Bennett}}, \bibinfo {author}
  {\bibfnamefont {M.~H.}\ \bibnamefont {Chen}}, \bibinfo {author}
  {\bibfnamefont {T.}~\bibnamefont {Cowan}}, \bibinfo {author} {\bibfnamefont
  {D.}~\bibnamefont {Dietrich}}, \bibinfo {author} {\bibfnamefont {J.~R.}\
  \bibnamefont {Henderson}}, \bibinfo {author} {\bibfnamefont {D.~A.}\
  \bibnamefont {Knapp}}, \bibinfo {author} {\bibfnamefont {A.}~\bibnamefont
  {Osterheld}}, \bibinfo {author} {\bibfnamefont {B.~M.}\ \bibnamefont
  {Penetrante}}, \bibinfo {author} {\bibfnamefont {M.~B.}\ \bibnamefont
  {Schneider}},\ and\ \bibinfo {author} {\bibfnamefont {J.~H.}\ \bibnamefont
  {Scofield}},\ }\bibfield  {title} {\bibinfo {title} {The use of an electron
  beam ion trap in the study of highly charged ions},\ }\href
  {https://doi.org/10.1016/0168-583X(89)90386-8} {\bibfield  {journal}
  {\bibinfo  {journal} {Nuclear Instruments and Methods in Physics Research
  Section B: Beam Interactions with Materials and Atoms}\ }\textbf {\bibinfo
  {volume} {43}},\ \bibinfo {pages} {431} (\bibinfo {year} {1989})}\BibitemShut
  {NoStop}%
\bibitem [{\citenamefont {{Crespo L\'opez-Urrutia}}\ \emph
  {et~al.}(1999{\natexlab{a}})\citenamefont {{Crespo L\'opez-Urrutia}},
  \citenamefont {Dorn}, \citenamefont {Moshammer},\ and\ \citenamefont
  {Ullrich}}]{crespo_1999a}%
  \BibitemOpen
  \bibfield  {author} {\bibinfo {author} {\bibfnamefont {J.~R.}\ \bibnamefont
  {{Crespo L\'opez-Urrutia}}}, \bibinfo {author} {\bibfnamefont
  {A.}~\bibnamefont {Dorn}}, \bibinfo {author} {\bibfnamefont {R.}~\bibnamefont
  {Moshammer}},\ and\ \bibinfo {author} {\bibfnamefont {J.}~\bibnamefont
  {Ullrich}},\ }\bibfield  {title} {\bibinfo {title} {The freiburg electron
  beam ion trap/source project {FreEBIT}},\ }\href
  {https://doi.org/10.1238/physica.topical.080a00502} {\bibfield  {journal}
  {\bibinfo  {journal} {Physica Scripta}\ }\textbf {\bibinfo {volume} {T80}},\
  \bibinfo {pages} {502} (\bibinfo {year} {1999}{\natexlab{a}})}\BibitemShut
  {NoStop}%
\bibitem [{\citenamefont {Dragani\ifmmode~\acute{c}\else \'{c}\fi{}}\ \emph
  {et~al.}(2003)\citenamefont {Dragani\ifmmode~\acute{c}\else \'{c}\fi{}},
  \citenamefont {{Crespo L\'opez-Urrutia}}, \citenamefont {DuBois},
  \citenamefont {Fritzsche}, \citenamefont {Shabaev}, \citenamefont {Orts},
  \citenamefont {Tupitsyn}, \citenamefont {Zou},\ and\ \citenamefont
  {Ullrich}}]{Draganic2003}%
  \BibitemOpen
  \bibfield  {author} {\bibinfo {author} {\bibfnamefont {I.}~\bibnamefont
  {Dragani\ifmmode~\acute{c}\else \'{c}\fi{}}}, \bibinfo {author}
  {\bibfnamefont {J.~R.}\ \bibnamefont {{Crespo L\'opez-Urrutia}}}, \bibinfo
  {author} {\bibfnamefont {R.}~\bibnamefont {DuBois}}, \bibinfo {author}
  {\bibfnamefont {S.}~\bibnamefont {Fritzsche}}, \bibinfo {author}
  {\bibfnamefont {V.~M.}\ \bibnamefont {Shabaev}}, \bibinfo {author}
  {\bibfnamefont {R.~S.}\ \bibnamefont {Orts}}, \bibinfo {author}
  {\bibfnamefont {I.~I.}\ \bibnamefont {Tupitsyn}}, \bibinfo {author}
  {\bibfnamefont {Y.}~\bibnamefont {Zou}},\ and\ \bibinfo {author}
  {\bibfnamefont {J.}~\bibnamefont {Ullrich}},\ }\bibfield  {title} {\bibinfo
  {title} {High precision wavelength measurements of {QED}-sensitive forbidden
  transitions in highly charged argon ions},\ }\href
  {https://doi.org/10.1103/PhysRevLett.91.183001} {\bibfield  {journal}
  {\bibinfo  {journal} {Phys. Rev. Lett.}\ }\textbf {\bibinfo {volume} {91}},\
  \bibinfo {pages} {183001} (\bibinfo {year} {2003})}\BibitemShut {NoStop}%
\bibitem [{\citenamefont {Orts}\ \emph {et~al.}(2006)\citenamefont {Orts},
  \citenamefont {Harman}, \citenamefont {L\'opez-Urrutia}, \citenamefont
  {Artemyev}, \citenamefont {Bruhns}, \citenamefont {Mart\'{\i}nez},
  \citenamefont {Jentschura}, \citenamefont {Keitel}, \citenamefont {Lapierre},
  \citenamefont {Mironov}, \citenamefont {Shabaev}, \citenamefont {Tawara},
  \citenamefont {Tupitsyn}, \citenamefont {Ullrich},\ and\ \citenamefont
  {Volotka}}]{SoriaOrts2006}%
  \BibitemOpen
  \bibfield  {author} {\bibinfo {author} {\bibfnamefont {R.~S.}\ \bibnamefont
  {Orts}}, \bibinfo {author} {\bibfnamefont {Z.}~\bibnamefont {Harman}},
  \bibinfo {author} {\bibfnamefont {J.~R.~C.}\ \bibnamefont {L\'opez-Urrutia}},
  \bibinfo {author} {\bibfnamefont {A.~N.}\ \bibnamefont {Artemyev}}, \bibinfo
  {author} {\bibfnamefont {H.}~\bibnamefont {Bruhns}}, \bibinfo {author}
  {\bibfnamefont {A.~J.~G.}\ \bibnamefont {Mart\'{\i}nez}}, \bibinfo {author}
  {\bibfnamefont {U.~D.}\ \bibnamefont {Jentschura}}, \bibinfo {author}
  {\bibfnamefont {C.~H.}\ \bibnamefont {Keitel}}, \bibinfo {author}
  {\bibfnamefont {A.}~\bibnamefont {Lapierre}}, \bibinfo {author}
  {\bibfnamefont {V.}~\bibnamefont {Mironov}}, \bibinfo {author} {\bibfnamefont
  {V.~M.}\ \bibnamefont {Shabaev}}, \bibinfo {author} {\bibfnamefont
  {H.}~\bibnamefont {Tawara}}, \bibinfo {author} {\bibfnamefont {I.~I.}\
  \bibnamefont {Tupitsyn}}, \bibinfo {author} {\bibfnamefont {J.}~\bibnamefont
  {Ullrich}},\ and\ \bibinfo {author} {\bibfnamefont {A.~V.}\ \bibnamefont
  {Volotka}},\ }\bibfield  {title} {\bibinfo {title} {Exploring relativistic
  many-body recoil effects in highly charged ions},\ }\href
  {https://doi.org/10.1103/PhysRevLett.97.103002} {\bibfield  {journal}
  {\bibinfo  {journal} {Phys. Rev. Lett.}\ }\textbf {\bibinfo {volume} {97}},\
  \bibinfo {pages} {103002} (\bibinfo {year} {2006})}\BibitemShut {NoStop}%
\bibitem [{\citenamefont {Bekker}\ \emph {et~al.}(2018)\citenamefont {Bekker},
  \citenamefont {Hensel}, \citenamefont {Daniel}, \citenamefont {Windberger},
  \citenamefont {Pfeifer},\ and\ \citenamefont {{Crespo
  L{\'o}pez-Urrutia}}}]{Bekker2018}%
  \BibitemOpen
  \bibfield  {author} {\bibinfo {author} {\bibfnamefont {H.}~\bibnamefont
  {Bekker}}, \bibinfo {author} {\bibfnamefont {C.}~\bibnamefont {Hensel}},
  \bibinfo {author} {\bibfnamefont {A.}~\bibnamefont {Daniel}}, \bibinfo
  {author} {\bibfnamefont {A.}~\bibnamefont {Windberger}}, \bibinfo {author}
  {\bibfnamefont {T.}~\bibnamefont {Pfeifer}},\ and\ \bibinfo {author}
  {\bibfnamefont {J.~R.}\ \bibnamefont {{Crespo L{\'o}pez-Urrutia}}},\
  }\bibfield  {title} {\bibinfo {title} {Laboratory precision measurements of
  optical emissions from coronal iron},\ }\href
  {https://doi.org/10.1103/PhysRevA.98.062514} {\bibfield  {journal} {\bibinfo
  {journal} {Phys. Rev. A}\ }\textbf {\bibinfo {volume} {98}},\ \bibinfo
  {pages} {062514} (\bibinfo {year} {2018})}\BibitemShut {NoStop}%
\bibitem [{\citenamefont {Bekker}\ \emph {et~al.}(2019)\citenamefont {Bekker},
  \citenamefont {Borschevsky}, \citenamefont {Harman}, \citenamefont {Keitel},
  \citenamefont {Pfeifer}, \citenamefont {Schmidt}, \citenamefont {Crespo
  L{\'o}pez-Urrutia},\ and\ \citenamefont {Berengut}}]{Bekker2019}%
  \BibitemOpen
  \bibfield  {author} {\bibinfo {author} {\bibfnamefont {H.}~\bibnamefont
  {Bekker}}, \bibinfo {author} {\bibfnamefont {A.}~\bibnamefont {Borschevsky}},
  \bibinfo {author} {\bibfnamefont {Z.}~\bibnamefont {Harman}}, \bibinfo
  {author} {\bibfnamefont {C.~H.}\ \bibnamefont {Keitel}}, \bibinfo {author}
  {\bibfnamefont {T.}~\bibnamefont {Pfeifer}}, \bibinfo {author} {\bibfnamefont
  {P.~O.}\ \bibnamefont {Schmidt}}, \bibinfo {author} {\bibfnamefont {J.~R.}\
  \bibnamefont {Crespo L{\'o}pez-Urrutia}},\ and\ \bibinfo {author}
  {\bibfnamefont {J.~C.}\ \bibnamefont {Berengut}},\ }\bibfield  {title}
  {\bibinfo {title} {Detection of the 5p -- 4f orbital crossing and its optical
  clock transition in {Pr$^{9+}$}},\ }\href
  {https://doi.org/10.1038/s41467-019-13406-9} {\bibfield  {journal} {\bibinfo
  {journal} {Nature Communications}\ }\textbf {\bibinfo {volume} {10}},\
  \bibinfo {pages} {5651} (\bibinfo {year} {2019})}\BibitemShut {NoStop}%
\bibitem [{\citenamefont {Rehbehn}\ \emph {et~al.}(2021)\citenamefont
  {Rehbehn}, \citenamefont {Rosner}, \citenamefont {Bekker}, \citenamefont
  {Berengut}, \citenamefont {Schmidt}, \citenamefont {King}, \citenamefont
  {Micke}, \citenamefont {Gu}, \citenamefont {M\"uller}, \citenamefont
  {Surzhykov},\ and\ \citenamefont {L\'opez-Urrutia}}]{Rehbehn_2021}%
  \BibitemOpen
  \bibfield  {author} {\bibinfo {author} {\bibfnamefont {N.-H.}\ \bibnamefont
  {Rehbehn}}, \bibinfo {author} {\bibfnamefont {M.~K.}\ \bibnamefont {Rosner}},
  \bibinfo {author} {\bibfnamefont {H.}~\bibnamefont {Bekker}}, \bibinfo
  {author} {\bibfnamefont {J.~C.}\ \bibnamefont {Berengut}}, \bibinfo {author}
  {\bibfnamefont {P.~O.}\ \bibnamefont {Schmidt}}, \bibinfo {author}
  {\bibfnamefont {S.~A.}\ \bibnamefont {King}}, \bibinfo {author}
  {\bibfnamefont {P.}~\bibnamefont {Micke}}, \bibinfo {author} {\bibfnamefont
  {M.~F.}\ \bibnamefont {Gu}}, \bibinfo {author} {\bibfnamefont
  {R.}~\bibnamefont {M\"uller}}, \bibinfo {author} {\bibfnamefont
  {A.}~\bibnamefont {Surzhykov}},\ and\ \bibinfo {author} {\bibfnamefont
  {J.~R.~C.}\ \bibnamefont {L\'opez-Urrutia}},\ }\bibfield  {title} {\bibinfo
  {title} {Sensitivity to new physics of isotope-shift studies using the
  coronal lines of highly charged calcium ions},\ }\href
  {https://doi.org/10.1103/PhysRevA.103.L040801} {\bibfield  {journal}
  {\bibinfo  {journal} {Phys. Rev. A}\ }\textbf {\bibinfo {volume} {103}},\
  \bibinfo {pages} {L040801} (\bibinfo {year} {2021})}\BibitemShut {NoStop}%
\bibitem [{\citenamefont {Gu}(2008)}]{gu_2008}%
  \BibitemOpen
  \bibfield  {author} {\bibinfo {author} {\bibfnamefont {M.~F.}\ \bibnamefont
  {Gu}},\ }\bibfield  {title} {\bibinfo {title} {The flexible atomic code},\
  }\href {https://doi.org/10.1139/p07-197} {\bibfield  {journal} {\bibinfo
  {journal} {Canadian Journal of Physics}\ }\textbf {\bibinfo {volume} {86}},\
  \bibinfo {pages} {675} (\bibinfo {year} {2008})}\BibitemShut {NoStop}%
\bibitem [{\citenamefont {{Crespo L\'opez-Urrutia}}\ \emph
  {et~al.}(1999{\natexlab{b}})\citenamefont {{Crespo L\'opez-Urrutia}},
  \citenamefont {Beiersdorfer}, \citenamefont {Widmann},\ and\ \citenamefont
  {Decaux}}]{crespo_1999b}%
  \BibitemOpen
  \bibfield  {author} {\bibinfo {author} {\bibfnamefont {J.~R.}\ \bibnamefont
  {{Crespo L\'opez-Urrutia}}}, \bibinfo {author} {\bibfnamefont
  {P.}~\bibnamefont {Beiersdorfer}}, \bibinfo {author} {\bibfnamefont
  {K.}~\bibnamefont {Widmann}},\ and\ \bibinfo {author} {\bibfnamefont
  {V.}~\bibnamefont {Decaux}},\ }\bibfield  {title} {\bibinfo {title} {Visible
  spectrum of highly charged ions: the forbidden optical lines of {Kr}, {Xe},
  and {Ba} ions in the {Ar I} to {Ni I} isoelectronic sequence},\ }\href
  {https://doi.org/10.1238/physica.topical.080a00448} {\bibfield  {journal}
  {\bibinfo  {journal} {Physica Scripta}\ }\textbf {\bibinfo {volume} {T80}},\
  \bibinfo {pages} {448} (\bibinfo {year} {1999}{\natexlab{b}})}\BibitemShut
  {NoStop}%
\bibitem [{\citenamefont {{Commission on Isotopic Abundances and Atomic Weights
  (CIAAW)}}(2003)}]{xenon_weight}%
  \BibitemOpen
  \bibfield  {author} {\bibinfo {author} {\bibnamefont {{Commission on Isotopic
  Abundances and Atomic Weights (CIAAW)}}},\ }\href
  {https://ciaaw.org/xenon.htm} {\bibinfo {title} {Atomic weight of xenon}}
  (\bibinfo {year} {2003}),\ \bibinfo {note} {(accessed:
  2021-08-15)}\BibitemShut {NoStop}%
\bibitem [{\citenamefont {Allehabi}\ \emph {et~al.}(2020)\citenamefont
  {Allehabi}, \citenamefont {Dzuba}, \citenamefont {Flambaum}, \citenamefont
  {Afanasjev},\ and\ \citenamefont {Agbemava}}]{Allehabi_2020}%
  \BibitemOpen
  \bibfield  {author} {\bibinfo {author} {\bibfnamefont {S.~O.}\ \bibnamefont
  {Allehabi}}, \bibinfo {author} {\bibfnamefont {V.~A.}\ \bibnamefont {Dzuba}},
  \bibinfo {author} {\bibfnamefont {V.~V.}\ \bibnamefont {Flambaum}}, \bibinfo
  {author} {\bibfnamefont {A.~V.}\ \bibnamefont {Afanasjev}},\ and\ \bibinfo
  {author} {\bibfnamefont {S.~E.}\ \bibnamefont {Agbemava}},\ }\bibfield
  {title} {\bibinfo {title} {Using isotope shift for testing nuclear theory:
  The case of nobelium isotopes},\ }\href
  {https://doi.org/10.1103/PhysRevC.102.024326} {\bibfield  {journal} {\bibinfo
   {journal} {Phys. Rev. C}\ }\textbf {\bibinfo {volume} {102}},\ \bibinfo
  {pages} {024326} (\bibinfo {year} {2020})}\BibitemShut {NoStop}%
\bibitem [{\citenamefont {Bloom}\ \emph {et~al.}(2014)\citenamefont {Bloom},
  \citenamefont {Nicholson}, \citenamefont {Williams}, \citenamefont
  {Campbell}, \citenamefont {Bishof}, \citenamefont {Zhang}, \citenamefont
  {Zhang}, \citenamefont {Bromley},\ and\ \citenamefont {Ye}}]{Bloom2014}%
  \BibitemOpen
  \bibfield  {author} {\bibinfo {author} {\bibfnamefont {B.~J.}\ \bibnamefont
  {Bloom}}, \bibinfo {author} {\bibfnamefont {T.~L.}\ \bibnamefont
  {Nicholson}}, \bibinfo {author} {\bibfnamefont {J.~R.}\ \bibnamefont
  {Williams}}, \bibinfo {author} {\bibfnamefont {S.~L.}\ \bibnamefont
  {Campbell}}, \bibinfo {author} {\bibfnamefont {M.}~\bibnamefont {Bishof}},
  \bibinfo {author} {\bibfnamefont {X.}~\bibnamefont {Zhang}}, \bibinfo
  {author} {\bibfnamefont {W.}~\bibnamefont {Zhang}}, \bibinfo {author}
  {\bibfnamefont {S.~L.}\ \bibnamefont {Bromley}},\ and\ \bibinfo {author}
  {\bibfnamefont {J.}~\bibnamefont {Ye}},\ }\bibfield  {title} {\bibinfo
  {title} {An optical lattice clock with accuracy and stability at the
  $10^{-18}$ level},\ }\href {https://doi.org/10.1038/nature12941} {\bibfield
  {journal} {\bibinfo  {journal} {Nature}\ }\textbf {\bibinfo {volume} {506}},\
  \bibinfo {pages} {71} (\bibinfo {year} {2014})}\BibitemShut {NoStop}%
\bibitem [{\citenamefont {Cing{\"o}z}\ \emph {et~al.}(2012)\citenamefont
  {Cing{\"o}z}, \citenamefont {Yost}, \citenamefont {Allison}, \citenamefont
  {Ruehl}, \citenamefont {Fermann}, \citenamefont {Hartl},\ and\ \citenamefont
  {Ye}}]{Cingoez_2012}%
  \BibitemOpen
  \bibfield  {author} {\bibinfo {author} {\bibfnamefont {A.}~\bibnamefont
  {Cing{\"o}z}}, \bibinfo {author} {\bibfnamefont {D.~C.}\ \bibnamefont
  {Yost}}, \bibinfo {author} {\bibfnamefont {T.~K.}\ \bibnamefont {Allison}},
  \bibinfo {author} {\bibfnamefont {A.}~\bibnamefont {Ruehl}}, \bibinfo
  {author} {\bibfnamefont {M.~E.}\ \bibnamefont {Fermann}}, \bibinfo {author}
  {\bibfnamefont {I.}~\bibnamefont {Hartl}},\ and\ \bibinfo {author}
  {\bibfnamefont {J.}~\bibnamefont {Ye}},\ }\bibfield  {title} {\bibinfo
  {title} {Direct frequency comb spectroscopy in the extreme ultraviolet},\
  }\href {https://doi.org/10.1038/nature10711} {\bibfield  {journal} {\bibinfo
  {journal} {Nature}\ }\textbf {\bibinfo {volume} {482}},\ \bibinfo {pages}
  {68} (\bibinfo {year} {2012})}\BibitemShut {NoStop}%
\bibitem [{\citenamefont {Lyu}\ \emph {et~al.}(2020)\citenamefont {Lyu},
  \citenamefont {Cavaletto}, \citenamefont {Keitel},\ and\ \citenamefont
  {Harman}}]{Lyu_2020}%
  \BibitemOpen
  \bibfield  {author} {\bibinfo {author} {\bibfnamefont {C.}~\bibnamefont
  {Lyu}}, \bibinfo {author} {\bibfnamefont {S.~M.}\ \bibnamefont {Cavaletto}},
  \bibinfo {author} {\bibfnamefont {C.~H.}\ \bibnamefont {Keitel}},\ and\
  \bibinfo {author} {\bibfnamefont {Z.}~\bibnamefont {Harman}},\ }\bibfield
  {title} {\bibinfo {title} {Interrogating the temporal coherence of {EUV}
  frequency combs with highly charged ions},\ }\href
  {https://doi.org/10.1103/PhysRevLett.125.093201} {\bibfield  {journal}
  {\bibinfo  {journal} {Phys. Rev. Lett.}\ }\textbf {\bibinfo {volume} {125}},\
  \bibinfo {pages} {093201} (\bibinfo {year} {2020})}\BibitemShut {NoStop}%
\bibitem [{\citenamefont {Nauta}\ \emph {et~al.}(2017)\citenamefont {Nauta},
  \citenamefont {Borodin}, \citenamefont {Ledwa}, \citenamefont {Stark},
  \citenamefont {Schwarz}, \citenamefont {Schmöger}, \citenamefont {Micke},
  \citenamefont {{Crespo López-Urrutia}},\ and\ \citenamefont
  {Pfeifer}}]{Nauta_2017}%
  \BibitemOpen
  \bibfield  {author} {\bibinfo {author} {\bibfnamefont {J.}~\bibnamefont
  {Nauta}}, \bibinfo {author} {\bibfnamefont {A.}~\bibnamefont {Borodin}},
  \bibinfo {author} {\bibfnamefont {H.~B.}\ \bibnamefont {Ledwa}}, \bibinfo
  {author} {\bibfnamefont {J.}~\bibnamefont {Stark}}, \bibinfo {author}
  {\bibfnamefont {M.}~\bibnamefont {Schwarz}}, \bibinfo {author} {\bibfnamefont
  {L.}~\bibnamefont {Schmöger}}, \bibinfo {author} {\bibfnamefont
  {P.}~\bibnamefont {Micke}}, \bibinfo {author} {\bibfnamefont {J.~R.}\
  \bibnamefont {{Crespo López-Urrutia}}},\ and\ \bibinfo {author}
  {\bibfnamefont {T.}~\bibnamefont {Pfeifer}},\ }\bibfield  {title} {\bibinfo
  {title} {Towards precision measurements on highly charged ions using a high
  harmonic generation frequency comb},\ }\href
  {https://doi.org/https://doi.org/10.1016/j.nimb.2017.04.077} {\bibfield
  {journal} {\bibinfo  {journal} {Nuclear Instruments and Methods in Physics
  Research Section B: Beam Interactions with Materials and Atoms}\ }\textbf
  {\bibinfo {volume} {408}},\ \bibinfo {pages} {285} (\bibinfo {year}
  {2017})},\ \bibinfo {note} {proceedings of the 18th International Conference
  on the Physics of Highly Charged Ions (HCI-2016), Kielce, Poland, 11-16
  September 2016}\BibitemShut {NoStop}%
\bibitem [{\citenamefont {Nauta}\ \emph {et~al.}(2021)\citenamefont {Nauta},
  \citenamefont {Oelmann}, \citenamefont {Borodin}, \citenamefont {Ackermann},
  \citenamefont {Knauer}, \citenamefont {Muhammad}, \citenamefont
  {Pappenberger}, \citenamefont {Pfeifer},\ and\ \citenamefont
  {L\'{o}pez-Urrutia}}]{Nauta_2021}%
  \BibitemOpen
  \bibfield  {author} {\bibinfo {author} {\bibfnamefont {J.}~\bibnamefont
  {Nauta}}, \bibinfo {author} {\bibfnamefont {J.-H.}\ \bibnamefont {Oelmann}},
  \bibinfo {author} {\bibfnamefont {A.}~\bibnamefont {Borodin}}, \bibinfo
  {author} {\bibfnamefont {A.}~\bibnamefont {Ackermann}}, \bibinfo {author}
  {\bibfnamefont {P.}~\bibnamefont {Knauer}}, \bibinfo {author} {\bibfnamefont
  {I.~S.}\ \bibnamefont {Muhammad}}, \bibinfo {author} {\bibfnamefont
  {R.}~\bibnamefont {Pappenberger}}, \bibinfo {author} {\bibfnamefont
  {T.}~\bibnamefont {Pfeifer}},\ and\ \bibinfo {author} {\bibfnamefont
  {J.~R.~C.}\ \bibnamefont {L\'{o}pez-Urrutia}},\ }\bibfield  {title} {\bibinfo
  {title} {{XUV} frequency comb production with an astigmatism-compensated
  enhancement cavity},\ }\href {https://doi.org/10.1364/OE.414987} {\bibfield
  {journal} {\bibinfo  {journal} {Opt. Express}\ }\textbf {\bibinfo {volume}
  {29}},\ \bibinfo {pages} {2624} (\bibinfo {year} {2021})}\BibitemShut
  {NoStop}%
\bibitem [{\citenamefont {Weyers}\ \emph {et~al.}(2018)\citenamefont {Weyers},
  \citenamefont {Gerginov}, \citenamefont {Kazda}, \citenamefont {Rahm},
  \citenamefont {Lipphardt}, \citenamefont {Dobrev},\ and\ \citenamefont
  {Gibble}}]{Weyers_2018}%
  \BibitemOpen
  \bibfield  {author} {\bibinfo {author} {\bibfnamefont {S.}~\bibnamefont
  {Weyers}}, \bibinfo {author} {\bibfnamefont {V.}~\bibnamefont {Gerginov}},
  \bibinfo {author} {\bibfnamefont {M.}~\bibnamefont {Kazda}}, \bibinfo
  {author} {\bibfnamefont {J.}~\bibnamefont {Rahm}}, \bibinfo {author}
  {\bibfnamefont {B.}~\bibnamefont {Lipphardt}}, \bibinfo {author}
  {\bibfnamefont {G.}~\bibnamefont {Dobrev}},\ and\ \bibinfo {author}
  {\bibfnamefont {K.}~\bibnamefont {Gibble}},\ }\bibfield  {title} {\bibinfo
  {title} {Advances in the accuracy, stability, and reliability of the {PTB}
  primary fountain clocks},\ }\href {https://doi.org/10.1088/1681-7575/aae008}
  {\bibfield  {journal} {\bibinfo  {journal} {Metrologia}\ }\textbf {\bibinfo
  {volume} {55}},\ \bibinfo {pages} {789} (\bibinfo {year} {2018})}\BibitemShut
  {NoStop}%
\bibitem [{\citenamefont {Gu{\'e}na}\ \emph {et~al.}(2017)\citenamefont
  {Gu{\'e}na}, \citenamefont {Weyers}, \citenamefont {Abgrall}, \citenamefont
  {Grebing}, \citenamefont {Gerginov}, \citenamefont {Rosenbusch},
  \citenamefont {Bize}, \citenamefont {Lipphardt}, \citenamefont {Denker},
  \citenamefont {Quintin}, \citenamefont {Raupach}, \citenamefont {Nicolodi},
  \citenamefont {Stefani}, \citenamefont {Chiodo}, \citenamefont {Koke},
  \citenamefont {Kuhl}, \citenamefont {Wiotte}, \citenamefont {Meynadier},
  \citenamefont {Camisard}, \citenamefont {{C Chardonnet}}, \citenamefont
  {Coq}, \citenamefont {Lours}, \citenamefont {Santarelli}, \citenamefont
  {{Amy-Klein}}, \citenamefont {Targat}, \citenamefont {Lopez}, \citenamefont
  {Pottie},\ and\ \citenamefont {Grosche}}]{guena_first_2017}%
  \BibitemOpen
  \bibfield  {author} {\bibinfo {author} {\bibfnamefont {J.}~\bibnamefont
  {Gu{\'e}na}}, \bibinfo {author} {\bibfnamefont {S.}~\bibnamefont {Weyers}},
  \bibinfo {author} {\bibfnamefont {M.}~\bibnamefont {Abgrall}}, \bibinfo
  {author} {\bibfnamefont {C.}~\bibnamefont {Grebing}}, \bibinfo {author}
  {\bibfnamefont {V.}~\bibnamefont {Gerginov}}, \bibinfo {author}
  {\bibfnamefont {P.}~\bibnamefont {Rosenbusch}}, \bibinfo {author}
  {\bibfnamefont {S.}~\bibnamefont {Bize}}, \bibinfo {author} {\bibfnamefont
  {B.}~\bibnamefont {Lipphardt}}, \bibinfo {author} {\bibfnamefont
  {H.}~\bibnamefont {Denker}}, \bibinfo {author} {\bibfnamefont
  {N.}~\bibnamefont {Quintin}}, \bibinfo {author} {\bibfnamefont {S.~M.~F.}\
  \bibnamefont {Raupach}}, \bibinfo {author} {\bibfnamefont {D.}~\bibnamefont
  {Nicolodi}}, \bibinfo {author} {\bibfnamefont {F.}~\bibnamefont {Stefani}},
  \bibinfo {author} {\bibfnamefont {N.}~\bibnamefont {Chiodo}}, \bibinfo
  {author} {\bibfnamefont {S.}~\bibnamefont {Koke}}, \bibinfo {author}
  {\bibfnamefont {A.}~\bibnamefont {Kuhl}}, \bibinfo {author} {\bibfnamefont
  {F.}~\bibnamefont {Wiotte}}, \bibinfo {author} {\bibfnamefont
  {F.}~\bibnamefont {Meynadier}}, \bibinfo {author} {\bibfnamefont
  {E.}~\bibnamefont {Camisard}}, \bibinfo {author} {\bibnamefont {{C
  Chardonnet}}}, \bibinfo {author} {\bibfnamefont {Y.~L.}\ \bibnamefont {Coq}},
  \bibinfo {author} {\bibfnamefont {M.}~\bibnamefont {Lours}}, \bibinfo
  {author} {\bibfnamefont {G.}~\bibnamefont {Santarelli}}, \bibinfo {author}
  {\bibfnamefont {A.}~\bibnamefont {{Amy-Klein}}}, \bibinfo {author}
  {\bibfnamefont {R.~L.}\ \bibnamefont {Targat}}, \bibinfo {author}
  {\bibfnamefont {O.}~\bibnamefont {Lopez}}, \bibinfo {author} {\bibfnamefont
  {P.~E.}\ \bibnamefont {Pottie}},\ and\ \bibinfo {author} {\bibfnamefont
  {G.}~\bibnamefont {Grosche}},\ }\bibfield  {title} {\bibinfo {title} {First
  international comparison of fountain primary frequency standards via a long
  distance optical fiber link},\ }\href
  {https://doi.org/10.1088/1681-7575/aa65fe} {\bibfield  {journal} {\bibinfo
  {journal} {Metrologia}\ }\textbf {\bibinfo {volume} {54}},\ \bibinfo {pages}
  {348} (\bibinfo {year} {2017})}\BibitemShut {NoStop}%
\bibitem [{\citenamefont {Rosenband}\ \emph {et~al.}(2008)\citenamefont
  {Rosenband}, \citenamefont {Hume}, \citenamefont {Schmidt}, \citenamefont
  {Chou}, \citenamefont {Brusch}, \citenamefont {Lorini}, \citenamefont
  {Oskay}, \citenamefont {Drullinger}, \citenamefont {Fortier}, \citenamefont
  {Stalnaker}, \citenamefont {Diddams}, \citenamefont {Swann}, \citenamefont
  {Newbury}, \citenamefont {Itano}, \citenamefont {Wineland},\ and\
  \citenamefont {Bergquist}}]{Rosenband2008}%
  \BibitemOpen
  \bibfield  {author} {\bibinfo {author} {\bibfnamefont {T.}~\bibnamefont
  {Rosenband}}, \bibinfo {author} {\bibfnamefont {D.~B.}\ \bibnamefont {Hume}},
  \bibinfo {author} {\bibfnamefont {P.~O.}\ \bibnamefont {Schmidt}}, \bibinfo
  {author} {\bibfnamefont {C.~W.}\ \bibnamefont {Chou}}, \bibinfo {author}
  {\bibfnamefont {A.}~\bibnamefont {Brusch}}, \bibinfo {author} {\bibfnamefont
  {L.}~\bibnamefont {Lorini}}, \bibinfo {author} {\bibfnamefont {W.~H.}\
  \bibnamefont {Oskay}}, \bibinfo {author} {\bibfnamefont {R.~E.}\ \bibnamefont
  {Drullinger}}, \bibinfo {author} {\bibfnamefont {T.~M.}\ \bibnamefont
  {Fortier}}, \bibinfo {author} {\bibfnamefont {J.~E.}\ \bibnamefont
  {Stalnaker}}, \bibinfo {author} {\bibfnamefont {S.~A.}\ \bibnamefont
  {Diddams}}, \bibinfo {author} {\bibfnamefont {W.~C.}\ \bibnamefont {Swann}},
  \bibinfo {author} {\bibfnamefont {N.~R.}\ \bibnamefont {Newbury}}, \bibinfo
  {author} {\bibfnamefont {W.~M.}\ \bibnamefont {Itano}}, \bibinfo {author}
  {\bibfnamefont {D.~J.}\ \bibnamefont {Wineland}},\ and\ \bibinfo {author}
  {\bibfnamefont {J.~C.}\ \bibnamefont {Bergquist}},\ }\bibfield  {title}
  {\bibinfo {title} {Frequency ratio of {Al+} and {Hg+} single-ion optical
  clocks; metrology at the 17th decimal place},\ }\href
  {https://doi.org/10.1126/science.1154622} {\bibfield  {journal} {\bibinfo
  {journal} {Science}\ }\textbf {\bibinfo {volume} {319}},\ \bibinfo {pages}
  {1808} (\bibinfo {year} {2008})}\BibitemShut {NoStop}%
\bibitem [{\citenamefont {Godun}\ \emph {et~al.}(2014)\citenamefont {Godun},
  \citenamefont {Nisbet-Jones}, \citenamefont {Jones}, \citenamefont {King},
  \citenamefont {Johnson}, \citenamefont {Margolis}, \citenamefont {Szymaniec},
  \citenamefont {Lea}, \citenamefont {Bongs},\ and\ \citenamefont
  {Gill}}]{Godun2014}%
  \BibitemOpen
  \bibfield  {author} {\bibinfo {author} {\bibfnamefont {R.~M.}\ \bibnamefont
  {Godun}}, \bibinfo {author} {\bibfnamefont {P.~B.~R.}\ \bibnamefont
  {Nisbet-Jones}}, \bibinfo {author} {\bibfnamefont {J.~M.}\ \bibnamefont
  {Jones}}, \bibinfo {author} {\bibfnamefont {S.~A.}\ \bibnamefont {King}},
  \bibinfo {author} {\bibfnamefont {L.~A.~M.}\ \bibnamefont {Johnson}},
  \bibinfo {author} {\bibfnamefont {H.~S.}\ \bibnamefont {Margolis}}, \bibinfo
  {author} {\bibfnamefont {K.}~\bibnamefont {Szymaniec}}, \bibinfo {author}
  {\bibfnamefont {S.~N.}\ \bibnamefont {Lea}}, \bibinfo {author} {\bibfnamefont
  {K.}~\bibnamefont {Bongs}},\ and\ \bibinfo {author} {\bibfnamefont
  {P.}~\bibnamefont {Gill}},\ }\bibfield  {title} {\bibinfo {title} {Frequency
  ratio of two optical clock transitions in {$^{171}{\mathrm{Yb}}^{+}$} and
  constraints on the time variation of fundamental constants},\ }\href
  {https://doi.org/10.1103/PhysRevLett.113.210801} {\bibfield  {journal}
  {\bibinfo  {journal} {Phys. Rev. Lett.}\ }\textbf {\bibinfo {volume} {113}},\
  \bibinfo {pages} {210801} (\bibinfo {year} {2014})}\BibitemShut {NoStop}%
\bibitem [{\citenamefont {Beloy}\ \emph {et~al.}(2021)\citenamefont {Beloy},
  \citenamefont {Bodine}, \citenamefont {Bothwell}, \citenamefont {Brewer},
  \citenamefont {Bromley}, \citenamefont {Chen}, \citenamefont {Desch{\^e}nes},
  \citenamefont {Diddams}, \citenamefont {Fasano}, \citenamefont {Fortier},
  \citenamefont {Hassan}, \citenamefont {Hume}, \citenamefont {Kedar},
  \citenamefont {Kennedy}, \citenamefont {Khader}, \citenamefont {Koepke},
  \citenamefont {Leibrandt}, \citenamefont {Leopardi}, \citenamefont {Ludlow},
  \citenamefont {McGrew}, \citenamefont {Milner}, \citenamefont {Newbury},
  \citenamefont {Nicolodi}, \citenamefont {Oelker}, \citenamefont {Parker},
  \citenamefont {Robinson}, \citenamefont {Romisch}, \citenamefont
  {Sch{\"a}ffer}, \citenamefont {Sherman}, \citenamefont {Sinclair},
  \citenamefont {Sonderhouse}, \citenamefont {Swann}, \citenamefont {Yao},
  \citenamefont {Ye}, \citenamefont {Zhang},\ and\ \citenamefont {{Boulder
  Atomic Clock Optical Network (BACON) Collaboration}}}]{Beloy2020}%
  \BibitemOpen
  \bibfield  {author} {\bibinfo {author} {\bibfnamefont {K.}~\bibnamefont
  {Beloy}}, \bibinfo {author} {\bibfnamefont {M.~I.}\ \bibnamefont {Bodine}},
  \bibinfo {author} {\bibfnamefont {T.}~\bibnamefont {Bothwell}}, \bibinfo
  {author} {\bibfnamefont {S.~M.}\ \bibnamefont {Brewer}}, \bibinfo {author}
  {\bibfnamefont {S.~L.}\ \bibnamefont {Bromley}}, \bibinfo {author}
  {\bibfnamefont {J.-S.}\ \bibnamefont {Chen}}, \bibinfo {author}
  {\bibfnamefont {J.-D.}\ \bibnamefont {Desch{\^e}nes}}, \bibinfo {author}
  {\bibfnamefont {S.~A.}\ \bibnamefont {Diddams}}, \bibinfo {author}
  {\bibfnamefont {R.~J.}\ \bibnamefont {Fasano}}, \bibinfo {author}
  {\bibfnamefont {T.~M.}\ \bibnamefont {Fortier}}, \bibinfo {author}
  {\bibfnamefont {Y.~S.}\ \bibnamefont {Hassan}}, \bibinfo {author}
  {\bibfnamefont {D.~B.}\ \bibnamefont {Hume}}, \bibinfo {author}
  {\bibfnamefont {D.}~\bibnamefont {Kedar}}, \bibinfo {author} {\bibfnamefont
  {C.~J.}\ \bibnamefont {Kennedy}}, \bibinfo {author} {\bibfnamefont
  {I.}~\bibnamefont {Khader}}, \bibinfo {author} {\bibfnamefont
  {A.}~\bibnamefont {Koepke}}, \bibinfo {author} {\bibfnamefont {D.~R.}\
  \bibnamefont {Leibrandt}}, \bibinfo {author} {\bibfnamefont {H.}~\bibnamefont
  {Leopardi}}, \bibinfo {author} {\bibfnamefont {A.~D.}\ \bibnamefont
  {Ludlow}}, \bibinfo {author} {\bibfnamefont {W.~F.}\ \bibnamefont {McGrew}},
  \bibinfo {author} {\bibfnamefont {W.~R.}\ \bibnamefont {Milner}}, \bibinfo
  {author} {\bibfnamefont {N.~R.}\ \bibnamefont {Newbury}}, \bibinfo {author}
  {\bibfnamefont {D.}~\bibnamefont {Nicolodi}}, \bibinfo {author}
  {\bibfnamefont {E.}~\bibnamefont {Oelker}}, \bibinfo {author} {\bibfnamefont
  {T.~E.}\ \bibnamefont {Parker}}, \bibinfo {author} {\bibfnamefont {J.~M.}\
  \bibnamefont {Robinson}}, \bibinfo {author} {\bibfnamefont {S.}~\bibnamefont
  {Romisch}}, \bibinfo {author} {\bibfnamefont {S.~A.}\ \bibnamefont
  {Sch{\"a}ffer}}, \bibinfo {author} {\bibfnamefont {J.~A.}\ \bibnamefont
  {Sherman}}, \bibinfo {author} {\bibfnamefont {L.~C.}\ \bibnamefont
  {Sinclair}}, \bibinfo {author} {\bibfnamefont {L.}~\bibnamefont
  {Sonderhouse}}, \bibinfo {author} {\bibfnamefont {W.~C.}\ \bibnamefont
  {Swann}}, \bibinfo {author} {\bibfnamefont {J.}~\bibnamefont {Yao}}, \bibinfo
  {author} {\bibfnamefont {J.}~\bibnamefont {Ye}}, \bibinfo {author}
  {\bibfnamefont {X.}~\bibnamefont {Zhang}},\ and\ \bibinfo {author}
  {\bibnamefont {{Boulder Atomic Clock Optical Network (BACON)
  Collaboration}}},\ }\bibfield  {title} {\bibinfo {title} {Frequency ratio
  measurements at 18-digit accuracy using an optical clock network},\ }\href
  {https://doi.org/10.1038/s41586-021-03253-4} {\bibfield  {journal} {\bibinfo
  {journal} {Nature}\ }\textbf {\bibinfo {volume} {591}},\ \bibinfo {pages}
  {564} (\bibinfo {year} {2021})}\BibitemShut {NoStop}%
\bibitem [{\citenamefont {Manovitz}\ \emph {et~al.}(2019)\citenamefont
  {Manovitz}, \citenamefont {Shaniv}, \citenamefont {Shapira}, \citenamefont
  {Ozeri},\ and\ \citenamefont {Akerman}}]{Manovitz2019}%
  \BibitemOpen
  \bibfield  {author} {\bibinfo {author} {\bibfnamefont {T.}~\bibnamefont
  {Manovitz}}, \bibinfo {author} {\bibfnamefont {R.}~\bibnamefont {Shaniv}},
  \bibinfo {author} {\bibfnamefont {Y.}~\bibnamefont {Shapira}}, \bibinfo
  {author} {\bibfnamefont {R.}~\bibnamefont {Ozeri}},\ and\ \bibinfo {author}
  {\bibfnamefont {N.}~\bibnamefont {Akerman}},\ }\bibfield  {title} {\bibinfo
  {title} {Precision measurement of atomic isotope shifts using a two-isotope
  entangled state},\ }\href {https://doi.org/10.1103/PhysRevLett.123.203001}
  {\bibfield  {journal} {\bibinfo  {journal} {Phys. Rev. Lett.}\ }\textbf
  {\bibinfo {volume} {123}},\ \bibinfo {pages} {203001} (\bibinfo {year}
  {2019})}\BibitemShut {NoStop}%
\bibitem [{\citenamefont {{Zyriliou, A.}}\ \emph {et~al.}(2022)\citenamefont
  {{Zyriliou, A.}}, \citenamefont {{Mertzimekis, T. J.}}, \citenamefont
  {{Chalil, A.}}, \citenamefont {{Vasileiou, P.}}, \citenamefont {{Mavrommatis,
  E.}}, \citenamefont {{Bonatsos, Dennis}}, \citenamefont {{Martinou,
  Andriana}}, \citenamefont {{Peroulis, S.}},\ and\ \citenamefont {{Minkov,
  N.}}}]{Zyriliou_2022}%
  \BibitemOpen
  \bibfield  {author} {\bibinfo {author} {\bibnamefont {{Zyriliou, A.}}},
  \bibinfo {author} {\bibnamefont {{Mertzimekis, T. J.}}}, \bibinfo {author}
  {\bibnamefont {{Chalil, A.}}}, \bibinfo {author} {\bibnamefont {{Vasileiou,
  P.}}}, \bibinfo {author} {\bibnamefont {{Mavrommatis, E.}}}, \bibinfo
  {author} {\bibnamefont {{Bonatsos, Dennis}}}, \bibinfo {author} {\bibnamefont
  {{Martinou, Andriana}}}, \bibinfo {author} {\bibnamefont {{Peroulis, S.}}},\
  and\ \bibinfo {author} {\bibnamefont {{Minkov, N.}}},\ }\bibfield  {title}
  {\bibinfo {title} {A study of some aspects of the nuclear structure in the
  even-even {Yb} isotopes},\ }\href
  {https://doi.org/10.1140/epjp/s13360-022-02414-2} {\bibfield  {journal}
  {\bibinfo  {journal} {Eur. Phys. J. Plus}\ }\textbf {\bibinfo {volume}
  {137}},\ \bibinfo {pages} {352} (\bibinfo {year} {2022})}\BibitemShut
  {NoStop}%
\end{thebibliography}%

%%% Supplemental Material is a separate document: main_supplemental_materials.tex

\end{document}

%\title{Sensitivity Study of Isotope Shifts to New Physics Fifth Forces using Optical Transitions of Highly Charged Xenon Ions}
\title{Supplemental Materials to:\\ Narrow and ultra-narrow transitions in highly charged Xe ions as probes of fifth forces}

\author{Nils-Holger Rehbehn\orcidA{}}
%\email[]{nils.rehbehn@mpi-hd.mpg.de}
\affiliation{Max-Planck-Institut f\"ur Kernphysik, D--69117 Heidelberg, Germany}

\author{Michael K. Rosner\orcidB{}}
\affiliation{Max-Planck-Institut f\"ur Kernphysik, D--69117 Heidelberg, Germany}

\author{Julian C. Berengut\orcidC{}}
\affiliation{School of Physics, University of New South Wales, Sydney, New South Wales 2052, Australia}
\affiliation{Max-Planck-Institut f\"ur Kernphysik, D--69117 Heidelberg, Germany}

\author{Piet O.~Schmidt\orcidD{}}
\affiliation{Physikalisch--Technische Bundesanstalt, D--38116 Braunschweig, Germany}
\affiliation{Leibniz Universit\"at Hannover, D--30167 Hannover, Germany}

\author{Thomas Pfeifer\orcidL{}}
\affiliation{Max-Planck-Institut f\"ur Kernphysik, D--69117 Heidelberg, Germany}

\author{Ming Feng Gu}
\affiliation{Space Science Laboratory, University of California, Berkeley, CA 94720, USA}

\author{Jos\'e R. {Crespo L\'opez-Urrutia}\orcidJ{}}
%\email[]{crespojr@mpi-hd.mpg.de}
\affiliation{Max-Planck-Institut f\"ur Kernphysik, D--69117 Heidelberg, Germany}

\date{\today}

\maketitle

\setcounter{figure}{0}
\renewcommand{\thefigure}{A\arabic{figure}} % change supplemental figures labeling
\setcounter{table}{0}
\renewcommand{\thetable}{B\arabic{table}} % change supplemental table labeling

\section{Other sensitive exclusion pairs}

\begin{figure*}[bh]
    \centering
    \includegraphics[width=\linewidth]{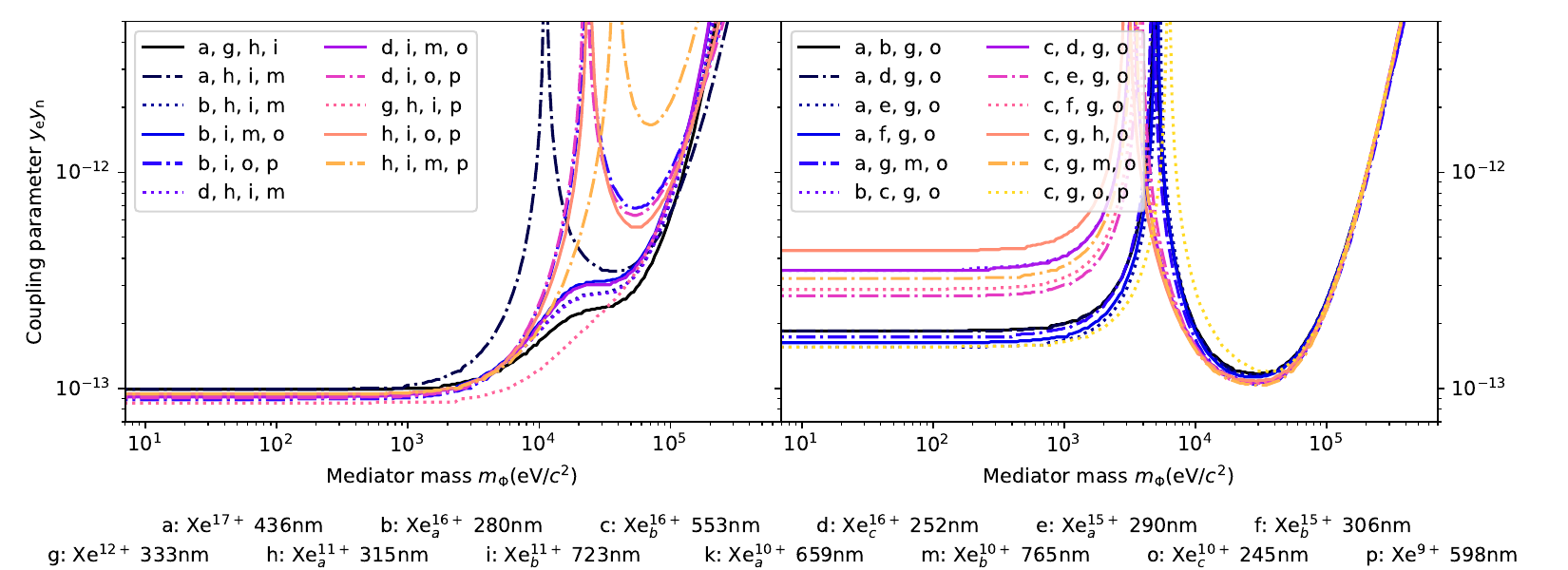}
    \caption{Lower limit to the coupling constant for different transition pairs to use in the no-mass generalized King plot yielding highest sensitivity. The letters indicate the transition, which can be found at the bottom of the figure.}
    \label{fig:Xe_mass_NMGKP_plot}
\end{figure*}

As the no-mass generalized King plot (GKP) in Fig. \ref{fig:Xe_exclusion} we only showed two possible pairs of transitions. Here, we wanted to mention other highly sensitive pairs.
Fig. \ref{fig:Xe_mass_NMGKP_plot} summarizes the transition pairs that yield the best possible sensitivity to a hypothetical fifth-force in the GKP. The order of the pairs do not matter, as the nonlinearity (NL) is the same. 
The pairs (g, h, i, p) and (a, g, m, o) were the ones shown in Fig. \ref{fig:Xe_exclusion}.
A listing of these transition-pairs can be found in Table \ref{tab:additional_exclusion_lines}, where the found lower limit of the coupling constant $y_\mathrm{e}y_\mathrm{n}$ is given for two mediator masses. 
Some of these pairs favor a light mediator mass, while others are more sensitive in the keV range. We separated those in the figure and in the table.
Other pairings were less sensitive and are omitted.

\begin{table*}
\centering
\caption{List of the exclusion lines from Fig. \ref{fig:Xe_mass_NMGKP_plot} with their corresponding lower limit of the coupling constant $y_\mathrm{e}y_\mathrm{n}$ for which the GKP method would be sensitive including current mass uncertainties.}
\begin{tabular}{llll|llll|r|r}
\hline
\hline
\multicolumn{8}{c|}{Transition pairing}& \multicolumn{2}{c}{$y_\mathrm{e}y_\mathrm{n}(m_\Phi)$} \\
\multicolumn{4}{c|}{Indices}&\multicolumn{4}{c|}{Transitions}& $m_\Phi=1~\mathrm{eV}/c^2$ & $m_\Phi=30~\mathrm{keV}/c^2$ \\
\hline
a & g & h & i & Xe$^{17+}$(436~nm), & Xe$^{12+}$(333~nm), & Xe$^{11+}$(315~nm), & Xe$^{11+}$(723~nm) & $9.9\times10^{-14}$ & $2.37\times10^{-13}$ \\
a & h & i & m & Xe$^{17+}$(436~nm), & Xe$^{11+}$(315~nm), & Xe$^{11+}$(723~nm), & Xe$^{10+}$(765~nm) & $9.9\times10^{-14}$ & $3.52\times10^{-13}$ \\
b & h & i & m & Xe$^{16+}$(280~nm), & Xe$^{11+}$(315~nm), & Xe$^{11+}$(723~nm), & Xe$^{10+}$(765~nm) & $9.1\times10^{-14}$ & $2.79\times10^{-13}$ \\
b & i & o & m & Xe$^{16+}$(280~nm), & Xe$^{11+}$(723~nm), & Xe$^{10+}$(765~nm), & Xe$^{10+}$(245~nm) & $9.2\times10^{-14}$ & $3.13\times10^{-13}$ \\
b & i & o & p & Xe$^{16+}$(280~nm), & Xe$^{11+}$(723~nm), & Xe$^{10+}$(245~nm), & Xe$^{ 9+}$(598~nm) & $8.8\times10^{-14}$ & $12.96\times10^{-13}$ \\
d & h & i & m & Xe$^{16+}$(252~nm), & Xe$^{11+}$(315~nm), & Xe$^{11+}$(723~nm), & Xe$^{10+}$(765~nm) & $9.1\times10^{-14}$ & $2.74\times10^{-13}$ \\
d & i & m & o & Xe$^{16+}$(252~nm), & Xe$^{11+}$(723~nm), & Xe$^{10+}$(765~nm), & Xe$^{10+}$(245~nm) & $9.1\times10^{-14}$ & $3.02\times10^{-13}$ \\
d & i & o & p & Xe$^{16+}$(252~nm), & Xe$^{11+}$(723~nm), & Xe$^{10+}$(245~nm), & Xe$^{ 9+}$(598~nm) & $9.0\times10^{-14}$ & $11.41\times10^{-13}$ \\
g & h & i & p & Xe$^{12+}$(333~nm), & Xe$^{11+}$(315~nm), & Xe$^{11+}$(723~nm), & Xe$^{ 9+}$(598~nm) & $8.6\times10^{-14}$ & $2.21\times10^{-13}$ \\
h & i & o & p & Xe$^{11+}$(315~nm), & Xe$^{11+}$(723~nm), & Xe$^{10+}$(245~nm), & Xe$^{ 9+}$(598~nm) & $9.4\times10^{-14}$ & $11.88\times10^{-13}$ \\
h & i & m & p & Xe$^{11+}$(315~nm), & Xe$^{11+}$(723~nm), & Xe$^{10+}$(765~nm), & Xe$^{ 9+}$(598~nm) & $9.4\times10^{-14}$ & $16.33\times10^{-13}$ \\
\hline
a & b & g & o & Xe$^{17+}$(436~nm), & Xe$^{16+}$(280~nm), & Xe$^{12+}$(333~nm), & Xe$^{ 10+}$(245~nm) & $18.4\times10^{-14}$ & $1.16\times10^{-13}$ \\
a & d & g & o & Xe$^{17+}$(436~nm), & Xe$^{16+}$(252~nm), & Xe$^{12+}$(333~nm), & Xe$^{ 10+}$(245~nm) & $18.4\times10^{-14}$ & $1.15\times10^{-13}$ \\
a & e & g & o & Xe$^{17+}$(436~nm), & Xe$^{15+}$(290~nm), & Xe$^{12+}$(333~nm), & Xe$^{ 10+}$(245~nm) & $15.5\times10^{-14}$ & $1.15\times10^{-13}$ \\
a & f & g & o & Xe$^{17+}$(436~nm), & Xe$^{15+}$(306~nm), & Xe$^{12+}$(333~nm), & Xe$^{ 10+}$(245~nm) & $16.3\times10^{-14}$ & $1.13\times10^{-13}$ \\
a & g & m & o & Xe$^{17+}$(436~nm), & Xe$^{12+}$(333~nm), & Xe$^{10+}$(765~nm), & Xe$^{ 10+}$(245~nm) & $17.3\times10^{-14}$ & $1.05\times10^{-13}$ \\
b & c & g & o & Xe$^{16+}$(280~nm), & Xe$^{16+}$(553~nm), & Xe$^{12+}$(333~nm), & Xe$^{ 10+}$(245~nm) & $35.2\times10^{-14}$ & $1.08\times10^{-13}$ \\
c & d & g & o & Xe$^{16+}$(553~nm), & Xe$^{16+}$(252~nm), & Xe$^{12+}$(333~nm), & Xe$^{ 10+}$(245~nm) & $35.2\times10^{-14}$ & $1.07\times10^{-13}$ \\
c & e & g & o & Xe$^{16+}$(553~nm), & Xe$^{15+}$(290~nm), & Xe$^{12+}$(333~nm), & Xe$^{ 10+}$(245~nm) & $26.8\times10^{-14}$ & $1.06\times10^{-13}$ \\
c & f & g & o & Xe$^{16+}$(553~nm), & Xe$^{15+}$(306~nm), & Xe$^{12+}$(333~nm), & Xe$^{ 10+}$(245~nm) & $28.7\times10^{-14}$ & $1.06\times10^{-13}$ \\
c & g & h & o & Xe$^{16+}$(553~nm), & Xe$^{12+}$(333~nm), & Xe$^{11+}$(315~nm), & Xe$^{ 10+}$(245~nm) & $43.4\times10^{-14}$ & $1.09\times10^{-13}$ \\
c & g & m & o & Xe$^{16+}$(553~nm), & Xe$^{12+}$(333~nm), & Xe$^{ 10+}$(765~nm), & Xe$^{10+}$(245~nm) & $32.3\times10^{-14}$ & $1.04\times10^{-13}$ \\
c & g & o & p & Xe$^{16+}$(553~nm), & Xe$^{12+}$(333~nm), & Xe$^{10+}$(245~nm), & Xe$^{ 9+}$(598~nm) & $15.5\times10^{-14}$ & $1.20\times10^{-13}$ \\

\hline
\hline
\end{tabular}
\label{tab:additional_exclusion_lines}
\end{table*}

\newpage
\section{Grotrian diagrams and other identified transitions}

\begin{figure*}
    \centering
    \includegraphics[width=\linewidth]{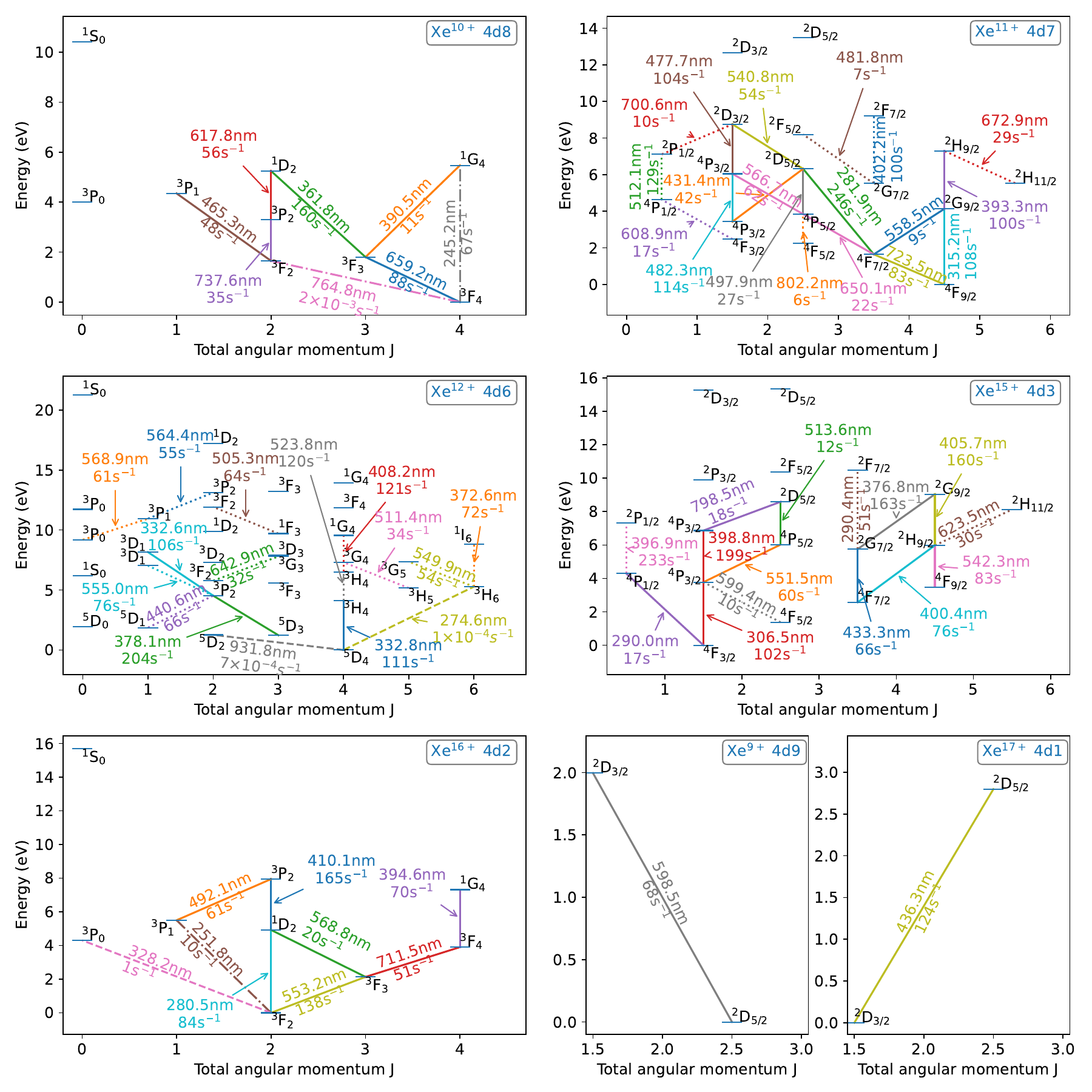}
    \caption{Grotrian diagram of Xe$^{9+}$ through Xe$^{17+}$. Solid lines are identified and measured transitions. Dotted lines are preliminary identifications. Dash-dotted lines are calculated results based on the identified transitions. Double-dashed lines are theoretical values based on calculations with \textsc{ambit}. Note the 764.8\,nm E2 transition in Xe$^{10+}$ inferred by the Ritz-Rydberg method. It arises from the lowest excited state, and with a calculated linewidth of 0.3\,mHz represents a suitable clock transition.}
    \label{fig:grotrian_all}
\end{figure*}

To summarize the identified transitions, besides the ones involving the ground-level, and to allow an insight about potentially interesting transitions, Grotrian diagrams are listed in Fig. \ref{fig:grotrian_all}. 
Xe$^{10+}$ has one E2 transition without an energy level in-between the upper and lower level, which can be found by calculating with the identified transitions.
Xe$^{12+}$ exhibits two potentially interesting E2 transitions, but further measurements are necessary to determine them and what energy the intermediate levels have. Their theoretically expected transitions and transition rates are added based on calculations by \textsc{ambit}.

\begin{table*}[bh]
\centering
\caption{Other identified transitions in Xe$^{10+}$. An E2 and a M1 transition were calculated based on identified transitions via Rydberg-Ritz combinations. Transition details, ab-initio wavelength $\lambda_\mathrm{ambit}$ and $A_{ki}$ values based on \textsc{ambit} calculations. 
}
\begin{tabular}{l|llll|l|l|r|r|l|l}
\hline
\hline
Ion & \multicolumn{4}{c|}{Transition} &	Energy (eV) & $\lambda_\mathrm{vac}$ (nm) & $\lambda_\mathrm{ambit}$ (nm) & $A_{ki}$ (s$^{-1}$) & Type & Note\\
\hline
Xe$^{10+}$&$4d^8$&$ ^1\!D_2 $&-&$ ^3\!\!F_3$ & 3.4276528(10) & 361.71748(10) & 353.6 & 159.5 &M1& \\ 
Xe$^{10+}$&$4d^8$&$ ^1\!G_4 $&-&$ ^3\!\!F_3$ & 3.1758602(72) & 390.39563(88) & 382.6 & 10.7 &M1& \\ 
Xe$^{10+}$&$4d^8$&$ ^3\!P_1 $&-&$ ^3\!\!F_2$ & 2.6655150(26) & 465.14163(45) & 453.7 & 47.8 &M1& \\ 
Xe$^{10+}$&$4d^8$&$ ^1\!D_2 $&-&$ ^3\!\!P_2$ & 2.0067271(28) & 617.84283(85) & 615.9 & 55.8 &M1& \\ 
Xe$^{10+}$&$4d^8$&$ ^3\!P_2 $&-&$ ^3\!\!F_2$ & 1.68081600(90) & 737.64289(39) & 741.4 & 35.3 &M1& \\ 
Xe$^{10+}$&$4d^8$&$ ^1\!G_4 $&-&$ ^3\!\!F_4$ & 5.0567231(73) & 245.18684(36) & 242.9 & 67.1 &M1& obtained by Ritz-Rydberg combination\\ 
Xe$^{10+}$&$4d^8$&$ ^3\!F_2 $&-&$ ^3\!\!F_4$ & 1.6209726(34) & 764.8753(16) & 735.6 & 0.002 &\textbf{E2}& obtained by Ritz-Rydberg combination\\ 
% Xe$^{10+}$&$4d^8$&$ ^3\!P_2 $&-&$ ^3\!\!F_4$ & 3.30195(16) & 375.487(18) & 0.18  &E2/M3& calculated, levels in between\\ 
% Xe$^{10+}$&$4d^8$&$ ^1\!D_2 $&-&$ ^3\!\!F_4$ & 5.30853(22) & 233.556(10) & 0.55  &E2/M3& calculated, levels in between\\ 
% Xe$^{10+}$&$4d^8$&$ ^3\!P_1 $&-&$ ^3\!\!F_4$ & 4.28605(46) & 289.273(31) & -  &M3& calculated, levels in between\\ 
\hline
\hline
\end{tabular}
\label{tab:extra_transitions_xenon10plus}
\end{table*}

Table \ref{tab:extra_transitions} lists the transitions from Fig. \ref{fig:grotrian_all} with their measured vacuum wavelength $\lambda_\mathrm{vac}$ and energy, as well as \textsc{ambit} ab-initio wavelengths and $A_{ki}$ values. Ground-level transitions from Table \ref{tab:results} are not listed. Ambiguous identifications are noted as preliminary and supplemental measurements would be needed. Interesting electric quadrupole E2 transitions were pointed out by noting them as \textsc{ambit} results. Where Rydberg-Ritz combination allowed identification, those were noted as such.

Xe$^{10+}$ has been separated to Table \ref{tab:extra_transitions_xenon10plus}, where an E2 transition was identified via Rydberg-Ritz combination. This clock transition has a lifetime of around 500s, thus a natural line-width of 0.3~mHz.

While Xe$^{12+}$ has two possible E2 transitions, no connection between measured lines was found and further investigation is required for an identification. It should be noted that we did not find the transitions to the intermediate levels, therefore it can not be ruled that they are energetically lower than the upper levels of the E2 transitions in this system.

The E2 transition in Xe$^{16+}$ might be possible to find via an infrared transition, however this requires further investigation as well. The needed transition between $^3P_0$ and $^3P_1$ is according to \textsc{ambit} at around 955~nm.

Rydberg-Ritz combination allowed us to find two M1 ground-level transitions in Xe$^{10+}$ and Xe$^{16+}$. Their wavelengths lie just outside the range of our conducted measurement.

\begin{table*}
\centering
\caption{Identified transitions in xenon, besides the ground-level transitions. Some interesting E2 transitions were calculated based on identified transitions. Some were added solely based on ab-initio \textsc{ambit} calculations, to point them out.
Lines with a preliminary note did not convey enough information to be identified without ambiguity, e.g., Zeeman structure was not sufficiently resolved or too many lines were too close to each other. Transition details, ab-initio wavelength $\lambda_\mathrm{ambit}$ and $A_{ki}$ values are based on \textsc{ambit} calculations. 
}
\begin{tabular}{l|llll|l|l|r|r|l|l}
% \begin{tabular}{p{0.05\linewidth}|llll|p{0.15\linewidth}|p{0.15\linewidth}|l|l|l}
\hline
\hline
Ion & \multicolumn{4}{c|}{Transition} &	Energy (eV) & $\lambda_\mathrm{vac}$ (nm) & $\lambda_\mathrm{ambit}$ (nm) & $A_{ki}$ (s$^{-1}$) & Type & Note\\
\hline
Xe$^{11+}$&$4d^7$&$ ^2\!D_{5/2} $&-&$ ^4\!\!F_{7/2}$  & 4.397932(52) & 281.9147(33) & 272.9 & 246.1 &M1& \\   
Xe$^{11+}$&$4d^7$&$ ^2\!H_{9/2} $&-&$ ^2\!\!G_{9/2}$  & 3.1523(48) & 393.3(6) & 395.9 & 99.6 &M1&  \\ 
Xe$^{11+}$&$4d^7$&$ ^2\!D_{5/2} $&-&$ ^4\!\!P_{3/2}$  & 2.8739(33) & 431.42(49) & 422.0 & 42.4 &M1& \\ 
Xe$^{11+}$&$4d^7$&$ ^2\!D_{3/2} $&-&$ ^4\!\!P_{3/2}$  & 2.5866(22) & 477.713(90) & 468.0 & 104.4 &M1& \\
Xe$^{11+}$&$4d^7$&$ ^4\!P_{3/2} $&-&$ ^4\!\!P_{3/2}$  & 2.5707010(33) & 482.29723(62) & 474.9 & 113.5 &M1& \\ 
Xe$^{11+}$&$4d^7$&$ ^2\!D_{5/2} $&-&$ ^4\!\!P_{5/2}$  & 2.4899(15) & 497.93(29) & 494.9 & 27.4 &M1& \\ 
Xe$^{11+}$&$4d^7$&$ ^2\!D_{3/2} $&-&$ ^2\!\!D_{5/2}$  & 2.29265(30) & 540.790(70) & 533.8 & 53.9 &M1& \\ 
Xe$^{11+}$&$4d^7$&$ ^2\!G_{9/2} $&-&$ ^4\!\!F_{7/2}$  & 2.21997(16) & 558.50(4) & 550.3 & 8.8 &M1& \\  
Xe$^{11+}$&$4d^7$&$ ^4\!P_{3/2} $&-&$ ^4\!\!P_{5/2}$  & 2.18799(39) & 566.66(10) & 569.2 & 61.8 &M1& \\ 
Xe$^{11+}$&$4d^7$&$ ^2\!P_{5/2} $&-&$ ^4\!\!F_{7/2}$  & 1.907216(88) & 650.079(30) & 608.4 & 16.7 &M1& \\
% Xe$^{11+}$&$4d^7$&$ ^2\!F_{7/2} $&-&$ ^2\!\!D_{5/2}$  & 2.59537(49) & 479.3(4) & & 3.2 &M1& preliminary \\ %transition seems to be in 13+ instead
Xe$^{11+}$&$4d^7$&$ ^2\!F_{7/2} $&-&$ ^2\!\!G_{7/2}$  & 3.0825(38) & 402.21(50) & 361.5 & 37.7 &M1& preliminary \\ 
Xe$^{11+}$&$4d^7$&$ ^2\!F_{5/2} $&-&$ ^2\!\!G_{7/2}$  & 2.5732(21) & 481.83(39) & 525.0 & 6.7 &M1& preliminary \\
Xe$^{11+}$&$4d^7$&$ ^2\!P_{1/2} $&-&$ ^4\!\!P_{1/2}$  & 2.42089(80) & 512.14(16) & 510.7 & 129.1 &M1& preliminary \\ 
Xe$^{11+}$&$4d^7$&$ ^4\!P_{1/2} $&-&$ ^4\!\!F_{3/2}$  & 2.03614(13) & 608.918(40) & 597.5 & 21.5 &M1& preliminary \\ 
Xe$^{11+}$&$4d^7$&$ ^2\!H_{9/2} $&-&$ ^2\!\!H_{11/2}$  & 1.842646(25) & 672.860(9) & 674.3 & 29.3 &M1& preliminary \\ 
Xe$^{11+}$&$4d^7$&$ ^2\!D_{3/2} $&-&$ ^2\!\!P_{1/2}$  & 1.76973(23) & 700.583(90) & 732.5 & 10.1 &M1& preliminary \\ 
Xe$^{11+}$&$4d^7$&$ ^4\!P_{5/2} $&-&$ ^4\!\!F_{5/2}$  & 1.54551(39) & 802.22(20) & 837.6 & 5.8 &M1& preliminary \\ 
\hline
Xe$^{12+}$&$4d^6$&$ ^3\!D_1 $&-&$ ^3\!\!P_2$ & 3.7282435(96) & 332.55391(86) & 340.1 & 105.5 &M1&  \\ 
Xe$^{12+}$&$4d^6$&$ ^3\!P_2 $&-&$ ^5\!\!D_3$ & 3.278867(10) & 378.1313(12) & 365.2 & 203.9 &M1&  \\  
Xe$^{12+}$&$4d^6$&$ ^1\!I_6 $&-&$ ^3\!\!H_6$ & 3.327889(18) & 372.5611(20) & 373.2 & 72.4 &M1& preliminary \\ 
Xe$^{12+}$&$4d^6$&$ ^1\!G_4 $&-&$ ^3\!\!H_4$ & 3.0372(30) & 408.21(39) & 394.5 & 121.3 &M1& preliminary \\  
Xe$^{12+}$&$4d^6$&$ ^3\!P_2 $&-&$ ^5\!\!D_1$ & 2.81415(10) & 440.57(15) & 459.0 & 66.0 &M1& preliminary \\  
Xe$^{12+}$&$4d^6$&$ ^3\!F_2 $&-&$ ^1\!\!F_3$ & 2.4539(11) & 505.25(22) & 480.6 & 63.6 &M1& preliminary \\  
Xe$^{12+}$&$4d^6$&$ ^3\!G_4 $&-&$ ^3\!\!H_5$ & 2.42449(90) & 511.38(18) & 505.6 & 33.8 &M1& preliminary \\  
Xe$^{12+}$&$4d^6$&$ ^3\!H_4 $&-&$ ^3\!\!H_4$ & 2.36708(45) & 523.79(10) & 527.9 & 120.3 &M1& preliminary \\  
Xe$^{12+}$&$4d^6$&$ ^3\!G_5 $&-&$ ^3\!\!H_6$ & 2.2546375(23) & 549.90744(55) & 540.0 & 54.3 &M1& preliminary \\ 
Xe$^{12+}$&$4d^6$&$ ^3\!D_1 $&-&$ ^3\!\!P_2$ & 2.23393(12) & 555.01(3) & 549.9 & 75.9 &M1& preliminary \\ 
Xe$^{12+}$&$4d^6$&$ ^3\!P_2 $&-&$ ^3\!\!P_1$ & 2.19676(47) & 564.39(11) & 584.9 & 60.8 &M1& preliminary \\ 
Xe$^{12+}$&$4d^6$&$ ^3\!P_1 $&-&$ ^3\!\!P_0$ & 2.17953(38) & 568.86(10) & 592.8 & 54.6 &M1& preliminary \\ 
Xe$^{12+}$&$4d^6$&$ ^3\!D_3 $&-&$ ^3\!\!F_2$ & 1.928492(90) & 642.907(30) & 636.9 & 31.6 &M1& preliminary \\ 
% Xe$^{12+}$&$4d^6$&$ ^5\!D_3 $&-&$ ^5\!\!D_4$ & 0.446370(14) & 2777.61(9) & & 1.5 &M1& calculated \\ 
Xe$^{12+}$&$4d^6$&$ ^5\!D_2 $&-&$ ^5\!\!D_4$& 1.3306 & & 931.8 & $7.2 \times 10^{-4}$ &\textbf{E2}& \textsc{ambit} result \\ 
Xe$^{12+}$&$4d^6$&$ ^3\!H_6 $&-&$ ^5\!\!D_4$& 4.5152 & & 274.6 & $9.6 \times 10^{-5}$ &\textbf{E2}& \textsc{ambit} result \\ 
% Xe$^{12+}$&$4d^6$&$ ^3\!P_2 $&-&$ ^5\!\!D_4$ & 4.174584(28) & 296.998(2) & & - &E2/M3& calculated, levels in between\\ 
\hline
Xe$^{15+}$&$4d^3$&$ ^2\!G_{9/2} $&-&$ ^2\!\!G_{7/2}$& 3.2906036(29) & 376.78253(33) & 376.7 & 162.6 &M1& \\ 
Xe$^{15+}$&$4d^3$&$ ^4\!P_{3/2} $&-&$ ^4\!\!P_{3/2}$& 3.1088401(49) & 398.81175(62) & 397.9 & 198.8 &M1& \\ 
Xe$^{15+}$&$4d^3$&$ ^2\!H_{9/2} $&-&$ ^4\!\!F_{7/2}$& 3.096290(15) & 400.428(2) & 394.2 & 76.5 &M1& \\ 
Xe$^{15+}$&$4d^3$&$ ^2\!G_{9/2} $&-&$ ^2\!\!H_{9/2}$& 3.056024(16) & 405.7042(21) & 400.6 & 160.5 &M1& \\
Xe$^{15+}$&$4d^3$&$ ^2\!G_{7/2} $&-&$ ^4\!\!F_{7/2}$& 2.861683(14) & 433.2561(21) & 420.4 & 65.7 &M1& \\ 
Xe$^{15+}$&$4d^3$&$ ^2\!D_{5/2} $&-&$ ^4\!\!P_{5/2}$& 2.41424(85) & 513.55(17) & 501.1 & 11.8 &M1&  \\
Xe$^{15+}$&$4d^3$&$ ^2\!H_{9/2} $&-&$ ^4\!\!F_{9/2}$& 2.286319(11) & 542.2875(25) & 536.6 & 82.9 &M1&  \\ 
Xe$^{15+}$&$4d^3$&$ ^4\!P_{5/2} $&-&$ ^4\!\!P_{3/2}$& 2.2483297(50) & 551.4502(12) & 554.1 & 59.6 &M1&  \\
Xe$^{15+}$&$4d^3$&$ ^2\!D_{5/2} $&-&$ ^3\!\!P_{3/2}$& 1.55279(97) & 798.45(49) & 777.0 & 18.1 &M1& \\ 
% Xe$^{15+}$&$4d^3$&$ ^2\!P_{3/2} $&-&$ ^2\!\!P_{1/2}$& 3.510534(58) & 353.1776(58) & & &M1& not identified \\ 
% Xe$^{15+}$&$4d^3$&$ ^2\!P_{3/2} $&-&$ ^2\!\!P_{1/2}$& 3.078083(38) & 402.7968(50) & & &M1& not identified \\  
% Xe$^{15+}$&$4d^3$&$ ^2\!P_{3/2} $&-&$ ^2\!\!P_{1/2}$& 2.8565688(43) & 434.03190(65) & & &M1& not identified \\
% Xe$^{15+}$&$4d^3$&$ ^2\!P_{3/2} $&-&$ ^2\!\!P_{1/2}$& 2.6111(27) & 474.83(49) & & &M1& not identified \\ 
% Xe$^{15+}$&$4d^3$&$ ^2\!P_{3/2} $&-&$ ^2\!\!P_{1/2}$& 2.6111(27) & 474.83(49) & & &M1& not identified \\ 
% Xe$^{15+}$&$4d^3$&$ ^2\!P_{3/2} $&-&$ ^2\!\!P_{1/2}$& 1.53613(38) & 807.12(19) & & &M1& not identified \\
Xe$^{15+}$&$4d^3$&$ ^2\!F_{7/2} $&-&$ ^2\!\!G_{7/2}$& 4.269411(14) & 290.40114(96) & 278.6 & 50.6 &M1& preliminary \\ 
Xe$^{15+}$&$4d^3$&$ ^2\!P_{1/2} $&-&$ ^4\!\!P_{1/2}$& 3.123792(52) & 396.9028(66) & 392.8 & 233.2 &M1& preliminary \\ 
Xe$^{15+}$&$4d^3$&$ ^4\!P_{3/2} $&-&$ ^4\!\!F_{5/2}$& 2.06849(17) & 599.395(50) & 548.0 & 10.3 &M1& preliminary \\ 
Xe$^{15+}$&$4d^3$&$ ^2\!H_{11/2} $&-&$ ^2\!\!H_{9/2}$& 1.988448(64) & 623.522(20) & 615.4 & 30.3 &M1& preliminary \\ 
\hline
Xe$^{16+}$&$4d^2$&$ ^1\!G_4 $&-&$ ^3\!\!F_4$& 3.14220(18) & 394.577(22) & 383.2 & 69.5 &M1&  \\ 
Xe$^{16+}$&$4d^2$&$ ^3\!P_2 $&-&$ ^1\!\!D_2$& 3.0230061(48) & 410.13545(65) & 407.7 & 165.2 &M1&  \\ 
Xe$^{16+}$&$4d^2$&$ ^3\!P_2 $&-&$ ^3\!\!P_1$& 2.5194098(20) & 492.11604(39) & 489.0 & 61.2 &M1&  \\ 
Xe$^{16+}$&$4d^2$&$ ^1\!D_2 $&-&$ ^3\!\!F_3$& 2.17991(38) & 568.8(1) & 530.3 & 20.2 &M1&  \\ 
Xe$^{16+}$&$4d^2$&$ ^3\!F_4 $&-&$ ^3\!\!F_3$& 1.7426542(51) & 711.4675(21) & 701.8 & 50.6 &M1&  \\ 
Xe$^{16+}$&$4d^2$&$ ^3\!P_1 $&-&$ ^3\!\!F_2$& 4.9243614(65) & 251.77721(33) & 244.3 & 10.3 &M1& obtained by Ritz-Rydberg combination \\ 
Xe$^{16+}$&$4d^2$&$ ^3\!P_0 $&-&$ ^3\!\!F_2$& 3.7766 & & 328.2 & 1.05 &\textbf{E2}& \textsc{ambit} result \\ 
% Xe$^{16+}$&$4d^2$&$ ^3\!F_4 $&-&$ ^3\!\!F_2$& 3.9836603(51) & 311.2319(4) & - &E2/M3& calculated, levels in between \\ 
\hline
\hline
\end{tabular}
\label{tab:extra_transitions}
\end{table*}